\documentclass{egpubl}
\usepackage{hpg25}
\CGFStandardLicense

\usepackage[T1]{fontenc}
\usepackage{dfadobe}
\usepackage{amsfonts}
\usepackage{amsmath}
\usepackage{subcaption}
\usepackage{algorithm}
\usepackage{algpseudocode}
\usepackage[table, dvipsnames]{xcolor}
\usepackage{color,soul}
\usepackage{caption}

\usepackage{cite}  
\BibtexOrBiblatex
\electronicVersion
\PrintedOrElectronic
\ifpdf \usepackage[pdftex]{graphicx} \pdfcompresslevel=9
\else \usepackage[dvips]{graphicx} \fi

\usepackage{breakurl}

\usepackage{rotating}
\usepackage{multirow}

\usepackage{egweblnk}
\title[DOBB-BVH]%
      {DOBB-BVH: Efficient Ray Traversal by Transforming Wide BVHs into Oriented Bounding Box Trees using Discrete Rotations}

\author[M. A. Kern \& A. Galvan \& D. R. Oldcorn \& D. Skinner \& R. Mehalwal \& L. Reyes Lozano \& M. G. Chajdas]
{\parbox{\textwidth}{\centering Michael A. Kern$^{1}$\orcid{0000-0002-8060-3367}
         Alain Galvan$^{1}$\orcid{0009-0000-2344-2993} David R. Oldcorn$^{1}$\hspace{2.5pt}Daniel Skinner$^{1}$\orcid{0009-0002-5837-8519} Rohan Mehalwal$^{1}$\orcid{0009-0006-1803-6855} Leo Reyes Lozano$^{1}$\hspace{2.5pt} \\Matth\"aus G. Chajdas$^{2}$\orcid{0000-0003-4689-2932}}
        \\
{\parbox{\textwidth}{\centering $^1$Advanced Micro Devices, Inc.\\
         $^2$Intel Corporation
}}
}

\begin{document}

\teaser{
  \includegraphics[width=\linewidth]{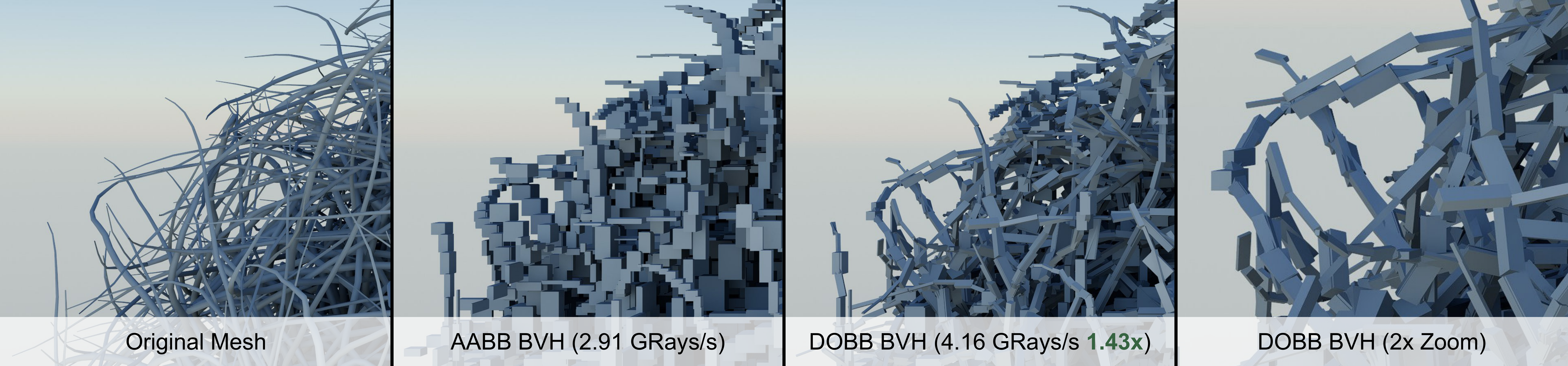}
  \centering
  \caption{Performance comparison and visualization of the lowest interior hierarchy nodes and their child bounding volumes in our DOBB-BVH, shown alongside its corresponding AABB-based BVH and the original triangle mesh.}
  \label{fig:teaser}
}

\maketitle
\begin{abstract}
{
Oriented bounding box (OBB) bounding volume hierarchies offer a more precise fit than axis-aligned bounding box hierarchies in scenarios with thin elongated and arbitrarily rotated geometry, enhancing intersection test performance in ray tracing. 
However, determining optimally oriented bounding boxes can be computationally expensive and have high memory requirements.
Recent research has shown that pre-built hierarchies can be efficiently converted to OBB hierarchies on the GPU in a bottom-up pass, yielding significant ray tracing traversal improvements. 
In this paper, we introduce a novel OBB construction technique where all internal node children share a consistent OBB transform, chosen from a fixed set of discrete quantized rotations. 
This allows for efficient encoding and reduces the computational complexity of OBB transformations. 
We further extend our approach to hierarchies with multiple children per node by leveraging Discrete Orientation Polytopes (\emph{k}-DOPs), demonstrating improvements in traversal performance while limiting the build time impact for real-time applications. Our method is applied as a post-processing step, integrating seamlessly into existing hierarchy construction pipelines. 
Despite a 12.6\% increase in build time, our experimental results demonstrate an average improvement of 18.5\% in primary, 32.4\% in secondary rays, and maximum gain of 65\% in ray intersection performance, highlighting its potential for advancing real-time applications.
}


\end{abstract}
\section{Introduction}
\label{sec:introduction}
\vspace*{-4pt}
In the computer graphics and gaming industry, recent advances in GPU hardware have enabled developers to synthesize photorealistic images and simulate light transport in real time. 
However, achieving this requires highly efficient intersection tests and occlusion queries. 
In ray tracing, bounding volume hierarchies (BVHs) with axis-aligned bounding boxes (AABBs) remain the predominant acceleration structure \cite{meister2021} to efficiently determine the closest or arbitrary intersections of light rays with scene geometry.
Although ray-AABB intersection tests are optimized in modern hardware \cite{rdna4_2025,blackwell_2025,alchemist_2024}, they often fail to tightly fit certain geometry, particularly thin elongated geometries not aligned with the three principal axes, such as hair strands or foliage. 
This inefficiency reduces the effectiveness of box culling and traversal performance.
%
To address this issue, prior research has explored alternative bounding volumes that provide tighter fits and improved hit detection performance.
These include spheres, oriented bounding boxes (OBBs), capsules, and discrete oriented polytopes with $k$ bounding planes (\emph{k}-DOPs)~\cite{schneider2003, ericson2004}. 
For elongated, arbitrarily rotated geometry, spheres tend to overestimate bounding volumes, while \emph{k}-DOPs introduce higher computational complexity and memory requirements. 
In contrast, orientation-aware bounding volumes, such as OBBs, offer a compelling alternative to AABBs, as they can closely fit arbitrarily oriented geometry with only a modest increase in computational cost and intersection complexity. 
Previous studies have shown that for structures like hair strands, cables, or branches following a dominant axis in 3D space, OBBs improve hit inference performance \cite{woop2014} in collision detection~\cite{rubin1980} or view frustum culling~\cite{gottschalk1996}. 

Nevertheless, the adoption of OBBs in real-time ray tracing applications has been hindered by three major challenges: (a) fast and robust OBB construction, (b) low memory representation of OBBs within hierarchical nodes, and (c) efficient ray-OBB intersection and traversal.
Recent research has made significant progress in fast and robust OBB construction \cite{vitsas2023, kacerik2024}, demonstrating that OBB bounding volume hierarchies (BVH) can be generated in milliseconds by converting AABB or \emph{k}-DOP hierarchies to OBBs as a post-processing step.
However, existing methods and studies rely on binary BVHs, do not impose constraints on BVH memory usage, and require non-SIMD-friendly ray-OBB intersection tests for each child node in the hierarchy.
In this paper, we extend prior work by addressing challenges (b) and (c) while also simplifying the construction of OBBs.
We present DOBB, a BVH that leverages discrete rotations and shared OBB transforms at the hierarchy node level to improve traversal performance in ray tracing.
Our approach is optimized for hardware-accelerated ray-OBB intersection tests.
As with prior methods, we employ a post-processing step to construct DOBB-BVHs from existing AABB hierarchies, enabling seamless integration into existing BVH formats and build pipelines.
The main contributions of this work are:
\begin{itemize}
    \item The introduction of a predefined set of discrete rotations using hemisphere and angle quantization.
    \item A novel OBB representation that maintains a common OBB transformation shared between all wide internal node children.
    \item An efficient orientation mapping technique using \emph{k}-DOP proxies interpreted as apex point maps~\cite{laine2015}.
    \item The usage of a compact encoding scheme that enables lossless reconstruction of OBB transformations.
    \item A comprehensive comparison between wide-node BVHs with shared orientations and per-child OBBs, evaluating BVH quality, traversal efficiency, and real-time performance on existing ray tracing hardware.
\end{itemize}
The remainder of this paper is structured as follows. Sect.~\ref{sec:related_work}, is concerned with related work. 
Sect.~\ref{sec:methodology} introduces the creation of discrete rotations, shared rotations, and our BVH build algorithm and traversal. 
Sect.~\ref{sec:implementation} details the implementation of our algorithm. 
In Sect.~\ref{sec:analysis}, we evaluate traversal and frame-time performance. 
Sect.~\ref{sec:conclusion} concludes this paper and discusses future work.

\vspace{-5pt}
\section{Related Work}
\label{sec:related_work}
\paragraph*{Bounding volume hierarchies}
Due to the latest advances in GPU technology and hardware-accelerated ray tracing, efficient parallel construction of BVHs on the GPU has been extensively surveyed~\cite{meister2021} in recent years.
Targeting GPU acceleration, linear bounding volume hierarchies~\cite{lauterbach2009, kerras2012, apetrei2014} are the most common approaches to rapidly construct BVHs.
Gu et al.~\cite{gu2013} first presented an agglomerative clustering technique for fast bottom-up BVH construction.
Karras and Aila~\cite{karras2013} extended this agglomerative approach and proposed to restructure small treelets.
Domingues and Pedrini~\cite{domingues2015} enhanced this approach by using a greedy clustering metric and widening binary BVHs while improving construction time with agglomerative treelet restructuring (ATRBVH).
Building on clustering, Meister and Bittner~\cite{meister2018} proposed a bottom-up clustering algorithm to construct BVHs while making use of spatial proximity of neighboring primitives, called parallel locally-ordered clustering (PLOC), and introduced a parallel topology reorganization algorithm~\cite{meister2018para}.
Benthin et al.~\cite{benthin2022} published PLOC++, a revised version of PLOC and with a highly improved construction time. 
More recently, Benthin et al.~\cite{benthin2024} also introduced H-PLOC, partitioning and clustering scene geometry guided by LBVH construction~\cite{apetrei2014} and invoking PLOC++ for nodes enclosing clusters with a target primitive count.
They showed that it competes with ATRBVH and PLOC++ and offers good BVH quality at fast BVH construction times.
Our paper uses H-PLOC to construct AABB BVHs, used as input to our OBB conversion step.

\vspace{-4pt}
\paragraph*{Orientation Computation}
Computing the optimal orientation or frame-of-reference for a given set of points or triangles has been the subject of past research since 1985.
The very first algorithm to compute optimal tight-fitting bounding boxes was presented by O'Rourke~\cite{orourke1985}.
A more popular approach is using principal component analysis (PCA)~\cite{gottschalk1996, klosowski1998}, where the predominant axes representing the largest variation in a point cloud set is determined in $\mathbb{R}^3$ space.
A drawback of PCA is that it introduces bias in the resulting orientation and thus fails to capture the orientation of enclosed geometry effectively in some edge cases~\cite{dimitrov2009}. 
Larsson and Linus~\cite{larsson2011} introduced the ditetrahedron OBB algorithm (DiTO) to tightly fit bounds in constant time. They analyzed a set of unique triangles, constructed a base triangle from extremal points, and constructed two tetrahedra facing in opposite directions. 
They have demonstrated that DiTO outperforms PCA in terms of robust OBB construction and can construct nearly optimal OBB rotations for arbitrary oriented geometry.

\begin{figure*}[!ht]
    \centering
    \includegraphics[width=1\linewidth]{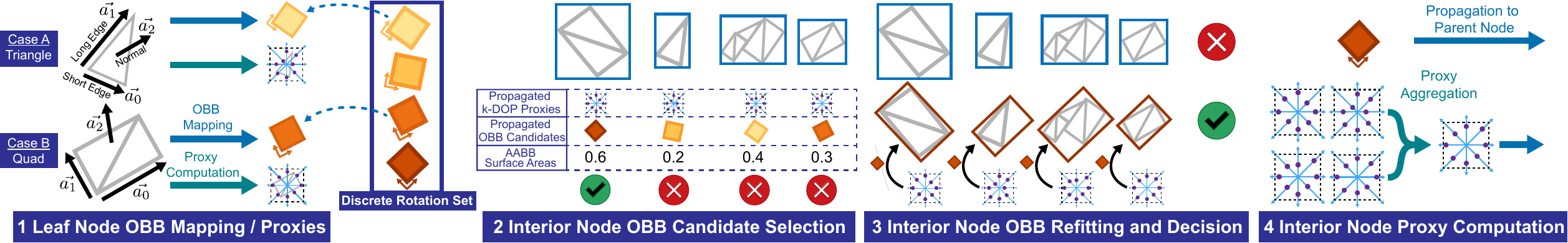}
    \caption{Overview of our OBB post-processing algorithm: 1) For each leaf node, the best rotation is derived from the triangle or quad frame and \emph{k}-DOP proxies are computed. 2) At interior nodes, the best OBB rotation candidate is selected based on the child node with the largest surface area. 3) Geometry is refit using the chosen rotation and proxies, and OBB data is stored if it improves over the AABB heuristic. 4) Proxies are combined and the OBB candidate is propagated up. Steps 2--4 repeat until a stopping criterion or the root node is reached.}
    \label{fig:obb-fitting}
\end{figure*}

\vspace{-4pt}
\paragraph*{OBB hierarchies}
The use of OBB hierarchies in ray tracing has gained increasing attention over the last decade. Woop et al.~\cite{woop2014} explored the effectiveness of OBBs for hair and fur rendering by leveraging the predominant orientation of long, thin primitives for intersection tests. They constructed a hybrid BVH using both AABBs and OBBs in a top-down manner within a CPU-based ray tracer.
Wald et al.~\cite{wald2020} utilized hardware-accelerated ray-AABB intersection tests and instance transformations to optimize ray tracing for elongated geometry. Their approach applied OBB-aware transformations at primitive nodes, demonstrating improved performance using modern APIs and GPUs.
More recently, Vitsas et al.~\cite{vitsas2023} introduced a method using \emph{k}-DOPs as proxies for sets of points per node, guiding the construction of DiTO to establish dominant directions for optimal rotations. Their algorithm converted AABBs into OBBs as a post-processing step on the GPU in a bottom-up manner, yielding significant ray tracing speedups for certain types of geometry. Sabino et al.~\cite{sabino2023} constructed OBBs from \emph{k}-DOPs with orthogonal bases (O-DOPs).
Building on this work, K\'{a}\v{c}erik and Bittner~\cite{kacerik2024} extended the use of \emph{k}-DOP proxies by integrating them into surface-area-aware BVH construction using PLOC. They demonstrated that these \emph{k}-DOPs could further improve OBB bounds, enhancing traversal efficiency and intersection performance.
More recently, K\'{a}\v{c}erik and Bittner introduced skewed OBBs and custom slab tests to improve ray traversal performance~\cite{kacerik2025}.
The work in this paper is based on the idea of using \emph{k}-DOP proxies and bottom-up OBB conversion. In contrast to previous research, however, our approach determines a consistent orientation for all children in an internal hierarchy node and uses much less memory to store this information while providing a SIMD-friendly ray-OBB intersection test.
Furthermore, our work is inspired by the DiTO approach since we face a similar challenge of finding optimal OBBs for arbitrarily oriented geometry.
As pointed out by Vitsas et al., the computational cost of exact construction algorithms is prohibitive for parallelization, while principal component analysis (PCA) suffers from limited accuracy.
In contrast to their approach, we further reduce computational complexity and memory lookups by restricting OBBs to a limited set of rotations.

\vspace{-8pt}
\section{Methodology}
\label{sec:methodology}
To enable the fast generation of our DOBB-BVH, we implemented a post-processing step in a bottom-up manner following the construction of an initial AABB BVH-N, where interior nodes contain up to $N$ children and leaf nodes enclose up to 2 triangles forming a quad.
Sect.~\ref{sec:implementation} covers the details of the wide AABB BVH creation.

A general overview of our designed GPU-driven post-processing approach is illustrated in Fig.~\ref{fig:obb-fitting}.
First, our algorithm computes the local frames of reference, leveraging topological information to efficiently select the optimal rotation from a discrete set of candidates.
Since our BVH leaf nodes are restricted to two triangles, we can rapidly compute face normals and obtain shared or non-shared edges to form an initial frame.
The core challenge of our work is then to identify best OBB candidates for $N$ child nodes for all internal hierarchy nodes. 
We begin by precomputing \emph{k}-DOPs for each leaf node in parallel. These proxies project the underlying geometry onto a limited, predefined set of axes and are stored as auxiliary sideband data during BVH construction, alongside the best-fitting OBB candidate.
This information is then propagated up the hierarchy to the parent interior nodes.
At each internal node, we use a heuristic-driven strategy to select consistent OBB rotation candidates from a fixed, discrete rotation set.
The OBB construction involves mapping proxies to the optimal orientation based on the propagated best-fitting OBB candidates and refitting the corresponding bounding volume extents using the underlying OBB rotation.
Once computed, we proceed with the final OBB encoding if the computed OBB bounding extents offer a measurable improvement in its cost heuristic.
For encoding, a compact representation using only a few bits is employed, alongside the AABBs defined in the rotated OBB space.
During bottom-up traversal, proxies of hierarchy nodes are updated at a higher level by merging \emph{k}-DOP proxies and forwarding the best OBB candidate information.
Our algorithm then repeats with identifying promising OBB candidates identification and refitting of child AABBs until it reaches the hierarchy root node or satisfies a stopping criterion, such as the maximum number of levels from the leaf node.

In the following sections, we detail the selection of discrete rotations and the steps involved in OBB construction. 
Finally, we describe the process of decoding OBBs and performing an efficient hardware-accelerated ray-OBB intersection.
\vspace{-8pt}
\subsection{Discrete Rotation Sets}
Unlike previous approaches, we defined a constrained set of discrete rotations for OBB construction. 
This is motivated by the observation that geometry within lower levels of a BVH tends to share similar orientations due to spatial proximity and topological similarity, making it more efficiently represented by OBBs.
Our goal is to define a rotation subset that balances flexibility and computational efficiency, yielding performance gains while ensuring fast build times. 
To achieve this, the number of possible rotations must be both controllable and precomputable for efficient decoding.
Throughout this paper, we denote an OBB rotation as $3\times3$ orthonormal matrix $R$, with $R R^T = I$.
Our final set of discrete rotations $\hat{R}_D$ comprises $D$ distinct rotations.
We constructed this set by quantizing rotation angles through uniform subdivision of the 2D circle, allowing control over the number of rotations around arbitrary axes.
Specifically, for a quantization step size of $\Delta = \frac{\pi}{2m}$ with up to $m$ angles in a single direction, our angle set is defined as:
\begin{equation*}
A:= \bigg\{\pm\Delta, \pm2\Delta, ..., \pm m\Delta\bigg\} \equiv \bigg\{\pm\frac{\pi}{2m}, \pm\frac{\pi}{m}, ..., \pm\frac{\pi}{2}\bigg\}\text{.}
\end{equation*}
For each axis, we selected angles in both positive and negative directions within the range $(0,\frac{\pi}{2}]$, excluding 0 to prevent identity rotations.
Empirically, we found that restricting rotations to $\pm\frac{\pi}{2}$ reduces redundant rotations around arbitrary axes compared to using the broader range $(0, \pi]$.

\vspace{-2pt}
Next, we defined a rotation representation that incorporates discrete angles, generating orthogonal, non-axis-aligned frames to construct our final set $\hat{R}_D$.
One common choice is the Euler angle representation, which parametrizes rotations using three quantized angles: yaw $\alpha$, pitch $\beta$, and roll $\gamma$. 
However, the Euler angle representation has several drawbacks, including: (a) \textbf{Gimbal lock}, where one rotation angle can effectively nullify the effect of the other two. (b) \textbf{Redundant rotations}, as different combinations of Euler angles can yield identical transformations. (c) \textbf{Higher computational and numerical costs}, since constructing the final rotation requires composing three matrices $R = R_z(\gamma) R_y(\beta) R_x(\alpha)$.
An alternative approach is to define an initial rotation axis using polar coordinates and apply quantized angles for the rotation around it.
This method requires storing both the selected quantized angle and the rotation axis, from which the final rotation can be reconstructed as either a quaternion or a $3\times3$ matrix.
While this approach avoids some of the pitfalls of Euler angles, it introduces additional complexity.
Specifically, the flexibility in choosing the initial axis makes it more difficult to control the total number of representable rotations and requires extra steps to reconstruct the final rotation matrix.

To overcome these challenges, we predefined a fixed set of rotation axes $K$ along with $A$.
This strategy offers two key advantages:
(a) It provides a approximately uniform distribution of axes across the 3D hemisphere, ensuring comprehensive coverage.
(b) It allows for efficient precomputation of a lookup table containing $D = |K| \cdot |A|$ discrete rotations, enabling fast retrieval and reducing computational overhead.
One possible way to create the set $K$ is to use a discrete representation of a sphere, such as Platonic solids.
Their well-defined vertex and face information allows us to create different combinations of unique axes along the topology. 
In combination with the angles in $A$, we eventually created table of discrete rotations.
In our work, we chose up to 13 axes and use Euclidean axes along with 6 diagonal axes lying within two planes and 4 axes in 3D space.
Notably, this method does not avoid duplicates in $R_D$ as some rotations inevitably produce similar sets of orthogonal vectors.
An example of how different configurations of $\hat{R}_D$ map to BVH quality is shown in Table~\ref{tab:discrete_sah_1} and discussed in Sect.~\ref{sec:analysis}.

\vspace*{-8pt}
\subsection{Per-Node Geometry Projections}
Our algorithm projects enclosed primitives onto \emph{k}-DOP axes at each leaf node to compute initial proxy bounds. 
This projection step is inherently parallel, as the leaf nodes reference independent primitives and can thus be processed in isolation.
For single-triangle leaves, we construct a local basis matrix $B = (\vec{a_0}, \vec{a_1}, \vec{a_2})$, where $\vec{a_0}$ and $\vec{a_1}$ correspond to the second-longest and shortest triangle edge, and $\vec{a_2} = \vec{a_0} \times \vec{a_1}$, as depicted in Fig.~\ref{fig:obb-fitting}.
Matrix $B$ is converted to an azimuth-angle representation and mapped to the closest orientation $R \in \hat{R}_D$ using a precomputed lookup table.
For quad leaves, we derive $B$ from the two non-shared edges, whereas for disconnected triangles, we default to the first triangle only.
The initial proxy data is propagated bottom-up through the hierarchy.
At each internal node, projection extents are aggregated along the same \emph{k}-DOP axes by merging the minimum and maximum values per axis from all children.
Parent nodes use this aggregated data during the OBB construction to compute the child OBB extents for a candidate rotation to eventually determine whether OBBs should be applied.
The projection values of our \emph{k}-DOP proxies are stored as sideband data along each internal node throughout the BVH construction.

\vspace*{-8pt}
\subsection{Consistent Orientation Determination}
A core step in our approach involves selecting the optimal rotation $R \in \hat{R}_D$ at each internal node to minimize the cost heuristic over its child bounding boxes.
Our algorithm begins by determining $R$ for leaf nodes and propagates the selected rotation upward through the BVH tree to rapidly determine candidate rotations.
While brute-force evaluation of all rotations yields optimal approach, it is impractical for real-time applications due to the high cost of recomputing bounds for all candidates and possible rotations.
An alternative is the DiTO algorithm, which constructs a base triangle, identifies extremal points, and tests multiple edges.
However, our experiments revealed that this method, while effective at hierarchy nodes higher up the tree, remains computationally expensive and does not consistently yield superior results compared to our approach.
Instead, we found empirically that using the propagated OBB candidate from the child with the largest surface area produces acceptable results at significantly lower cost.
This reduces candidate selection to a single rotation and requires only one reprojection of child node extents.
With this, wave intrinsics can be used to efficiently identify the largest-area child.
If the chosen $R$ reduces the cost heuristic, we construct an OBB at the internal node and replace the AABB bounds with rotated equivalents for all children.

\vspace{-8pt}
\subsection{Bounds Recomputation}
\begin{figure}[!h]
  \centering
  \includegraphics[width=1\linewidth]{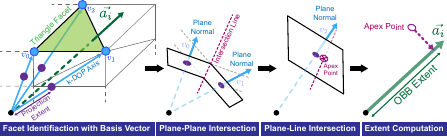}
  \caption{Schematic of OBB extent computation. The rotation basis vector $a_i$ traverses a triangle facet of a tessellated quad-sphere, whose vertices $v_i$ rerpresent a \emph{k}-DOP axis. Planes are constructed using those axes and the projection extent. Their intersection yields the apex point, which is projected onto $a_i$ to obtain the final extents.}
  \label{fig:obb_extent}
\end{figure}
Once the candidate OBB rotation $R$ is determined, we compute the bounding planes of all interior node children in rotated space.
Contrary to~\cite{vitsas2023}, which recomputes the axial projection extents for \emph{k}-DOPs at each internal node using all primitive vertices within the subtree, we avoid this costly operation due to its high memory access overhead.
Instead, we restrict vertex projections to leaf nodes only and utilize \emph{k}-DOP proxies to determine the OBB extents of interior node children.
However, since the OBB basis vectors induced by $R$ are generally misaligned with the \emph{k}-DOP axes, a direct use of our proxies is not possible. This misalignment requires a conservative computation of the three bounding planes.
To address this, we employ apex point maps, which enable fast and conservative fitting of bounding planes for arbitrarily oriented geometries. 
The technique operates on a tessellated surface of triangle facets, in our work a triangulated quad-sphere, where the facet vertices define directions corresponding to the proxy axes.
The three OBB extents for each child node are computed using this representation, as illustrated in Fig.~\ref{fig:obb_extent}.
%
The algorithm starts by converting $R$ to its three orthonormal basis vectors $\vec{a_0}$, $\vec{a_1}$, and $\vec{a_2}$.
These vectors are tested against the quad-sphere.
Once the intersecting triangle facet with vertices $v_0$, $v_1$, and $v_2$ is identified for each $\vec{a_i}$, the apex points are computed using geometric intersection tests.
Specifically, we define planes using the \emph{k}-DOP axes corresponding to $v_0$ and $v_1$ as normals and the projection extent as distance and perform a plane-plane intersection, followed by a plane-line intersection with the plane associated with $v_2$ to determine the final apex point.
Final OBB extents are computed from the three apex points by projecting those onto $\vec{a_i}$ via a dot product.

\vspace*{-8pt}
\subsection{OBB Cost Heuristic}
Since bounding boxes are no longer constrained to align with the Euclidean axes, a key challenge lies in defining an effective cost heuristic to guide the creation of OBB nodes.
We focused on the surface area heuristic (SAH), widely used in BVH research and designed to guide BVH construction such that the computed cost approximates the expected ray tracing performance, see~\cite{aila2013}.
To keep our algorithm computationally efficient while maintaining effectiveness, we require a heuristic that minimizes SAH and is both lightweight and impactful.
Potential options include: (a) surface area minimization, (b) child-to-child overlap reduction, and (c) approximate \emph{k}-DOP SAH computation~\cite{kacerik2024}.
Approach (c) involves sampling the BVH in a top-down manner, explicitly computing the SAH for \emph{k}-DOPs to assess whether an OBB reduces intersection likelihood. However, this method is computationally expensive, requires evaluation at every node and offers limited parallelism. Additionally, their approach is tailored to 14-DOPs, whereas our approach uses 26-DOPs, enabling tighter OBB construction.
Approach (b) requires $(N-1)!$ computations for a node with $N$ children and does not necessarily improve traversal performance unless normalized and combined with surface area minimization, which introduces an empirically determined weighting factor.
In our experiments, minimizing the total surface area of all child bounding boxes proved to be the most effective and computationally efficient strategy, consistent with the assumption that smaller bounds lower intersection probability for arbitrary rays.

Defining an internal node $N$ with $n$ child nodes $C_i$ and the surface area $SA_R$ of their bounding boxes defined in the transformed space $R$, we finally determine the node cost:
\begin{equation}
    C_R(N) = \sum_{i = 0}^n SA_R(C_i).
    \label{eq:cost}
\end{equation}
We construct an OBB with rotation $R$ whenever $C_R(N) < \alpha~C_I(N)$, where $I$ is the reference identity matrix and AABB frame, and $\alpha$ is a user-defined threshold.
Although we recommend choosing $\alpha$ from the range $(0.9, 1]$, larger values $> 1$ can be used to soften the OBB creation criterion.
In our experiments, we found that this could also improve performance depending on the scene and cast rays.

\vspace{-8pt}
\subsection{OBB Encoding Scheme}

Keeping memory usage as low as possible is essential for real-time applications, necessitating a compact BVH node layout. 
Given that all child node AABBs are defined in a common rotated space, we encode $R$ within the internal node data using only a few additional bits.
To facilitate both efficient reconstruction and minimal memory overhead, we store all rotations in $\hat{R}_D$ using two precomputed lookup tables.
The first table stores a pool of shared floating-point components reused across all $3\times3$ rotation matrices. The second table specifies each rotation $R \in \hat{R_D}$ by indexing into the first, effectively encoding each matrix using 12 references to shared values.
Since our OBB construction is limited to $D$ rotations, each internal node only requires at most $\left\lceil(log_2(D)\right\rceil$) additional bits.
During BVH traversal, decoding is efficient: the stored index is used to retrieve the corresponding matrix via the second table, with final floating-point values reconstructed by referencing the first table.

\subsection{BVH Traversal}
\begin{algorithm}[!h]
\caption{Our SIMD GPU kernel to traverse DOBB-BVHs.}
\begin{algorithmic}
\label{alg:traversal}
\State $S \gets Stack([\text{root}])$
\State ray $\gets$ origin,direction
\State $H \gets$ invalidNode,ray.tMax
\While{not $S$.IsEmpty()}
    \State node $\gets$ FetchNode(S.PopNextHitCandidate())
    \If{IsInteriorNode(node)}
        \State $R \gets$ FetchTransform(node.obbIndex)
        \State localRay $\gets$ RayTransform(ray, $R$)
        \State hits $\gets$ IntersectRayNodeChildren(localRay, node)
        \State S.SortAndPushCandidates(hits)
    \ElsIf{IsTriangleNode(node)}
        \State hits $\gets$ IntersectRayTriangles(node)
        \State H.Update(hits)
    \EndIf
\EndWhile
\end{algorithmic}
\end{algorithm}
Our BVH traversal is based on a SIMD stack-based traversal loop executed on the GPU. At each iteration, we decode the node’s AABBs, child offsets, and, if present, the OBB transform. 
The ray is then tested for intersection against all child nodes. 
The traversal stack maintains the next hit candidates, ordered by a heuristic that prioritizes the objects most likely to be hit.
For OBB-based BVHs, the local rotation matrix $R$ is obtained as previously described. The input ray is transformed into the rotated space prior to intersection with all internal node children. 
Although reusing the transformed ray across nodes with identical rotations is theoretically possible, changes in $R$ require applying the inverse of the previous transformation, potentially introducing numerical errors through repeated transformations. To mitigate this, our implementation consistently recomputes the local ray directly from the original one, at the cost of using additional vector registers to store the local ray.
A key advantage of this approach is compatibility with existing hardware-accelerated ray-box intersection intrinsics, given that the BVH is stored in the appropriate hardware memory layout. 
Additionally, we leverage hardware-accelerated ray-matrix transformations to enable ray-OBB intersection.
Regarding memory latency, fetching a node index requires a single memory access to load the node from VRAM or cache, while retrieving $R$ involves an additional lookup in constant global memory.

\vspace{-8pt}
\section{Implementation}
\label{sec:implementation}
To generate wide AABB-BVHs, we first build our BVH-2 using H-PLOC~\cite{benthin2024} offering a balance between fast build times and quality.
Previous studies have shown that SAH-optimal widening of BVH-2 significantly reduces traversal iterations by enabling more effective splits~\cite{alia2009}, improving performance on GPUs due to its SIMD-friendly nature.
Therefore, a widening step is applied in a top-down manner to expand the BVH-2 nodes to our target width~\cite{ylitie2017} of 8 children, while optimizing for SAH.
Since our leaf nodes may contain up to two triangles, we aim to form quads. 
Quadification can be performed either as a pre-processing step by forming triangle pairs while minimizing the surface area of their bounding box, or during the BVH widening phase by merging two leaf nodes enclosed by a single interior node.

Based on our studies, we opted for $\hat{R}_{104}$ as the discrete rotations set for our final performance studies with $|K| = 13$ axes and $|A| = 8$ angles ($m = 4$).
This allows us to encode an OBB index with just 7 bits at each interior node.
The algorithm is implemented as a single HLSL kernel based on shader model 6.3 in Direct3D12 that executes after the construction of a BVH-N and functions as a bottom-up post-process pass.
It starts with an initial list of internal nodes that have only leaves as children, and use $N$ threads per element in this list, where $N$ is the number of children per internal node.
Every internal node is then processed with up to $N$ threads to pick the child with largest surface area and its best $R \in \hat{R}_{104}$.
To track and recompute child node bounds, we use \emph{26}-DOPs stored along with each node.
After refitting of bounds and computing the cost heuristic, our kernel encodes the final OBB index and bounds if Eq.~\ref{eq:cost} is met.
Refitting fits within a work-group processor and local cache while utilizing a single fetch instruction to speed up execution.
In our traversal kernel, we make use of a single HW intrinsic performing both ray transform and intersection based on the encoded OBB index and bounding boxes for an $N$ wide interior hierarchy node.

In addition to our GPU-based approach, we developed a CPU-based reference version of our algorithm to estimate the theoretical upper bound of potential performance improvements achievable with DOBB-BVHs. 
Unlike the GPU variant, this version performs a brute-force evaluation of all possible OBB candidates. 
For each candidate rotation, it computes projection extents along all unique $225$ basis vectors derived from the full set of rotations in $\hat{R}_{104}$. 
This approach eliminates the need of apex point maps and enables exact computation of bounding box extents in the rotated space.
Rather than selecting the OBB candidate from the child node with the largest surface area, the CPU algorithm evaluates the complete set of rotations, computes the corresponding bounding volumes, and selects $R$ that minimizes the cost function.

All our kernels and following performance measurements were run on an AMD Radeon RX 9070 XT GPU with 16GB GDDR6 video memory and boost frequency up to 2.9 GHz. 
CPU BVH builds were executed on an AMD Ryzen 9 5950X processor.
\vspace{-8pt}
\section{Analysis}
\label{sec:analysis}

We evaluated our approach in terms of traversal behavior, memory consumption, build time, and real-time performance.
To compare our GPU algorithm to speed-of-light BVH builds and OBB construction, we make use of our offline CPU BVH post-processing step as explained in Sect.~\ref{sec:implementation}, applied to the pregenerated AABB BVH-Ns.
We further use a branching factor $N \in {4, 6, 8}$ in our studies to demonstrate the benefit of our algorithm for wide BVHs.
In the remainder of this section, we refer to OBB$_\text{PC}$ with the construction of OBBs for each single child encoded in sideband data alongside all hierarchy nodes.
OBB$_\text{soCPU}$ denotes the CPU-based method, while OBB$_\text{soGPU}$ represents our GPU-based approach.

\begin{table}[!t]
  \centering
  \small
  \begin{tabular}{p{3pt}r|c|c|c|c|c|c|c|c|}
    \multicolumn{2}{l}{} & \multicolumn{8}{c}{Total Number of Rotation Angles and Discretization Steps ($m$)} \\
    & & \cellcolor{darkgray}\textcolor{white}{2(1)} & \cellcolor{darkgray}\textcolor{white}{4(2)} & \cellcolor{darkgray}\textcolor{white}{6(3)} & \cellcolor{darkgray}\textcolor{white}{8(4) } & \cellcolor{darkgray}\textcolor{white}{10(5)} & \cellcolor{darkgray}\textcolor{white}{12(6)} & \cellcolor{darkgray}\textcolor{white}{14(7)} & \cellcolor{darkgray}\textcolor{white}{16(8)} \\ \cline{2-10}
    \multirow{11}{*}{\rotatebox[origin=c]{90}{Total Number of Rotation Axes}} & \cellcolor{darkgray}\textcolor{white}{3} & \textcolor{purple}{195.6} & 167.2 & 158.0 & 144.6 & 141.8 & 149.5 & 148.3 & 147.5 \\ 
    & \cellcolor{darkgray}\textcolor{white}{4} & 195.6 & 167.2 & 141.9 & 132.6 & 140.8 & 140.0 & 141.5 & 147.5 \\ 
    & \cellcolor{darkgray}\textcolor{white}{5} & 173.9 & 148.8 & 133.3 & 127.3 & 130.3 & 137.6 & 138.6 & 137.7 \\ 
    & \cellcolor{darkgray}\textcolor{white}{6} & 173.9 & 148.8 & 131.2 & 124.5 & 124.9 & 128.7 & 135.0 & 137.7 \\ 
    & \cellcolor{darkgray}\textcolor{white}{7} & 173.9 & 141.6 & 128.8 & 122.6 & 122.6 & 124.5 & 127.6 & 133.0 \\ 
    & \cellcolor{darkgray}\textcolor{white}{8} & 173.9 & 141.6 & 125.9 & 120.6 & 120.8 & 122.2 & 124.6 & 126.9 \\ 
    & \cellcolor{darkgray}\textcolor{white}{9} & 163.6 & 136.5 & 123.4 & 119.2 & 119.7 & 120.7 & 122.0 & 124.7 \\ 
    & \cellcolor{darkgray}\textcolor{white}{10} & 163.6 & 136.5 & 123.2 & 118.4 & 118.2 & 119.0 & 120.2 & 121.6 \\ 
    & \cellcolor{darkgray}\textcolor{white}{11} & 163.6 & 132.2 & 122.7 & 117.5 & \textcolor{PineGreen}{\textbf{116.9}} & 117.9 & 118.9 & 119.8 \\ 
    & \cellcolor{darkgray}\textcolor{white}{12} & 163.6 & 132.2 & 121.8 & \textcolor{PineGreen}{\textbf{116.7}} & \textcolor{PineGreen}{\textbf{116.2}} & 116.9 & 117.6 & 119.0 \\ 
    & \cellcolor{darkgray}\textcolor{white}{13} & 153.5 & 128.9 & 121.3 & \textcolor{PineGreen}{\textbf{116.0}} & \textcolor{PineGreen}{\textbf{115.6}} & 115.7 & 116.5 & 117.6 \\ 
    \cline{3-10}
  \end{tabular}
  \caption{SAH computed on DOBB-based BVH-8 generated from \emph{Hairball} with OBB$_\text{soGPU}$ using different axes and angles counts, highlighting \textcolor{purple}{axis-aligned orientations only} and \textcolor{PineGreen}{promising configs}.}
  \label{tab:discrete_sah_1}
\end{table}

\vspace{-8pt}
\subsection{Datasets}
For our studies, we selected 5 scenes with different levels of complexity, as shown in Table~\ref{tab:data_sah_iters}. \emph{Hairball} is a scene with hair-like geometry and randomly oriented hair strands well suited for OBBs and posing a challenge for AABB BVHs to well represent the underlying geometry. The \emph{White Oak} is a tree with triangulated leaves of different orientation in 3D space. 
\emph{Dragon} represents a mixture of axis and non-axis aligned geometry where parts of the model, such as the sword and tail, will benefit from OBBs particularly well. 
\emph{Bistro Exterior} scene is comprised of the same amount of triangles as hairball but with a mixture of axis-aligned and rotated buildings, furniture, and foliage.
\emph{Powerplant} contains massive triangle geometry and some non-axis aligned geometry, such as construction crossbars that can lead to highly increased iteration counts due to multi-intersection caused by multiple ill-fitting AABBs.

\vspace{-8pt}
\subsection{BVH Quality}
We first evaluated the quality of BVHs by measuring SAH. Prior research has shown that SAH approximately correlates with the final ray tracing performance for rays cast from the outside of the geometry~\cite{aila2013}. 
For both OBB approaches, we compute an approximate SAH by taking the child AABB stored in the parent node as base to compute the SAH of its subtree.
To determine an appropriate configuration for $\hat{R}_D$, we evaluated various discrete orientation sets $R_D$ with $|A| \in [3,13]$ and $m \in [1,8]$, where $|K| = 2m$, with consistently oriented hair strands. 
We measured the resulting SAH values of DOBB-BVHs constructed for the \emph{Hairball} scene using a branching factor of $N=8$ (see Table~\ref{tab:discrete_sah_1}).
As expected, the SAH reduction improves with an increased number of discrete rotations. 
However, the improvements diminish beyond 10 unique axes.
Notably, a configuration with $m = 1$ and all three Euclidean axes can be neglected as a 90-degree rotation simply inverts the axes.

Based on this analysis, we selected $D = 13 \cdot 8 = 104$ discrete rotations as a balanced trade-off between construction quality and memory overhead.
Using $\hat{R}_{104}$, we then evaluated SAH values across all five datasets and multiple BVH branching factors (refer to Table~\ref{tab:data_sah_iters}). The most significant SAH reductions occur in the \emph{Hairball} and \emph{Bistro Exterior} scenes, where OBB$_\text{soCPU}$ achieves a 40--47\% SAH reduction compared to AABB BVH-Ns, consistently across all branching factors. In comparison, OBB$_\text{soGPU}$ reduces SAH by approximately 20–25\%.
The \emph{Powerplant} dataset benefits similarly from oriented bounding volumes. 
OBB$_\text{soGPU}$ achieves a 15--17\% SAH reduction, only about 10\% below the results of the CPU-based offline builder.
Conversely, the \emph{White Oak} and \emph{Dragon} scenes reveal limitations of shared OBBs. 
These scenes feature characteristics such as small primitives, highly axis-aligned geometry, or randomized orientations, which makes it challenging to identify fitting OBBs. 
In these cases, SAH reduction is limited to 14\%, with most results in the 2--6\% range. 
This trend worsens with increasing branching factors, highlighting the challenge of coherent orientation fitting in wide BVHs.
Overall, OBB$_\text{soCPU}$ delivers performance comparable to per-child orientation approach used in OBB$_\text{PC}$, particularly in scenes with strong orientation coherence.

\vspace{-8pt}
\subsection{Traversal Behavior}

We analyzed the effect of BVH quality on traversal efficiency by measuring per-ray iteration steps needed to find the closest triangle hit. On our target GPU, which supports HW-accelerated ray intersection with 8 OBB child nodes and 2 triangles per step, iteration count corresponds directly to node visits. Since long-running rays can stall entire SIMD wavefronts, we evaluated both maximum and average iteration counts (see Table~\ref{tab:data_sah_iters}) and visualized traversal cost per pixel for BVH-8 with primary and global illumination (GI) rays (Tables~\ref{tab:bvh8_iters} and \ref{tab:bvh8_iters_gi}, App.~\ref{app:analysis}).
The most notable gains were observed in maximum iteration counts. For \emph{Hairball}, \emph{Powerplant}, and \emph{Dragon}, OBB$_\text{soCPU}$ reduced max. iterations by 40–80\%, and OBB$_\text{soGPU}$ by 30--40\%. In \emph{Dragon}, even minor SAH improvements helped addressing poorly approximated tail regions. Similarly, in \emph{Powerplant}, rays traversing non-axis-aligned crossbars caused over 1000 iterations, significantly mitigated by OBBs. GI rays showed even greater improvements, with max. iteration counts up to 3$\times×$ higher than primary rays.
OBB$_\text{soCPU}$ reduced these by up to 91\% for \emph{Powerplant} and 82\% for \emph{Hairball} (Table~\ref{tab:data_sah_iters_gi}).
We found that average iteration reductions are less pronounced. For primary rays, OBB$_\text{soCPU}$ achieved 20–25\% improvements, and OBB$_\text{soGPU}$ 17--22\%. Axis-aligned scenes like \emph{White Oak} saw limited gains (1--8\%), as orientation fitting is less effective. This gets increasingly challenging for wider BVHs, where a consistent orientation becomes harder to identify. Therefore, OBB$_\text{PC}$ generally outperforms shared-orientation DOBBs. 
Still, OBB$_\text{soCPU}$ delivers comparable results, improving average traversal by 10--20\% over OBB$_\text{soGPU}$.

\begin{table}[t!]
    \centering
    \small
    \begin{tabular}{|l|l|l|l|l|}
        \hline 
        \cellcolor{darkgray} & \cellcolor{darkgray} & \multicolumn{1}{c|}{\cellcolor{darkgray}\textcolor{white}{Build Time}} & \multicolumn{1}{c|}{\cellcolor{darkgray}\textcolor{white}{Primary}} & \multicolumn{1}{c|}{\cellcolor{darkgray}\textcolor{white}{Secondary}} \\
    \multirow{-2}{*}{\cellcolor{darkgray}\textcolor{white}{Scene}} & \multirow{-2}{*}{\cellcolor{darkgray}\textcolor{white}{Approach}} & \multicolumn{1}{c|}{\cellcolor{darkgray}\textcolor{white}{ms}} & \multicolumn{1}{c|}{\cellcolor{darkgray}\textcolor{white}{GRays/s}} & \multicolumn{1}{c|}{\cellcolor{darkgray}\textcolor{white}{GRays/s}} \\ \hline
        ~ & AABB & 11.62 & 2.91 & 1.16 \\ 
        ~ & OBB$_\text{soGPU}$ & 12.82 \tiny\textcolor{purple}{(1.10$\times$)} & \textbf{4.16 }\textcolor{teal}{\tiny(\textbf{1.43$\times$})} & \textbf{1.66} \tiny\textcolor{teal}{(\textbf{1.43$\times$})} \\
        \multirow{-3}{*}{Hairball} & OBB$_\text{soCPU}$ & 
        \textcolor{gray}{\textit{-}} 
        & {5.10} \textcolor{teal}{\tiny(\textbf{1.76$\times$})} & {2.16} \tiny\textcolor{teal}{({1.86$\times$})} \\ \hline
        ~ & AABB & 12.16 &  3.63 & 1.34 \\ 
        ~ & OBB$_\text{soGPU}$ & 13.56 \tiny\textcolor{purple}{(1.12$\times$)} &  4.27 \tiny\textcolor{teal}{(1.18$\times$)} &  \textbf{1.97} \tiny\textcolor{teal}{(\textbf{1.47$\times$})} \\
        \multirow{-3}{*}{Bistro Ext.} & OBB$_\text{soCPU}$ & 
        \textcolor{gray}{\textit{-}} 
        & 5.01 \tiny\textcolor{teal}{(1.38$\times$)} &  {2.77} \tiny\textcolor{teal}{({2.06$\times$})} \\ \hline
        ~ & AABB & 0.65 & 7.95 & 2.82 \\ 
        ~ & OBB$_\text{soGPU}$ & 0.71 \tiny\textcolor{purple}{(1.09$\times$)} & 8.04 \tiny\textcolor{teal}{(1.01$\times$)} & 2.93 \tiny\textcolor{teal}{(1.04$\times$)} \\
        \multirow{-3}{*}{White Oak} & OBB$_\text{soCPU}$ & 
        \textcolor{gray}{\textit{-}} 
        & 8.23 \tiny\textcolor{teal}{(1.04$\times$)} & 3.06 \tiny\textcolor{teal}{(1.08$\times$)} \\ \hline
        ~ & AABB & 2.57 & 20.93 & 3.26 \\ 
        ~ & OBB$_\text{soGPU}$ & 3.02 \tiny\textcolor{purple}{(1.18$\times$)} &  21.08 \tiny\textcolor{teal}{(1.01$\times$)} &  3.37 \tiny\textcolor{teal}{(1.03$\times$)} \\
        \multirow{-3}{*}{Dragon} & OBB$_\text{soCPU}$ & 
        \textcolor{gray}{\textit{-}} 
        &  20.87 \tiny\textcolor{teal}{(1.00$\times$)} &  3.47 \tiny\textcolor{teal}{(1.06$\times$)} \\ \hline
        ~ & AABB & 51.02 &  4.09 & 1.87 \\ 
        ~ & OBB$_\text{soGPU}$ & 55.20 \tiny\textcolor{purple}{(1.08$\times$)} & \textbf{5.32} \tiny\textcolor{teal}{(\textbf{1.30$\times$})} & \textbf{3.09} \tiny\textcolor{teal}{(\textbf{1.65$\times$})} \\
         \multirow{-3}{*}{Powerplant} & OBB$_\text{soCPU}$ &
         \textcolor{gray}{\textit{-}} 
         & {4.96} \tiny\textcolor{teal}{({1.21$\times$})} & {3.68} \tiny\textcolor{teal}{({1.96$\times$})} \\ \hline
    \end{tabular}
    \caption{Performance of our CPU- and GPU-based DOBB BVH-8 build algorithm compared against an AABB tree and the relative factor within brackets \small(\textcolor{purple}{increased build time} or \textcolor{teal}{improved performance}).}
    \label{tab:hw_results_bvh8}
\end{table}

\begin{table*}[!h]
  \centering
  \small
  \renewcommand{\arraystretch}{0.99}
  \begin{tabular}{c|c|c||c|c|c||c|c|c||c|c|c|}
    \cline{2-12}
    & \cellcolor{black}\textcolor{white}{Scene} & \cellcolor{black}\textcolor{white}{Approach} & \multicolumn{3}{c||}{\cellcolor{black}\textcolor{white}{Surface Area Heuristic (SAH)}} & \multicolumn{3}{c||}{\cellcolor{black}\textcolor{white}{Max. Iterations / Ray}} & \multicolumn{3}{c|}{\cellcolor{black}\textcolor{white}{Avg. Iterations / Ray}}\\
    & \cellcolor{darkgray}\textcolor{white}{\tiny [\#Triangles]} & & \cellcolor{darkgray}\textcolor{white}{BVH-4} & \cellcolor{darkgray}\textcolor{white}{BVH-6} & \cellcolor{darkgray}\textcolor{white}{BVH-8} & \cellcolor{darkgray}\textcolor{white}{BVH-4} & \cellcolor{darkgray}\textcolor{white}{BVH-6} & \cellcolor{darkgray}\textcolor{white}{BVH-8} & \cellcolor{darkgray}\textcolor{white}{BVH-4} & \cellcolor{darkgray}\textcolor{white}{BVH-6} & \cellcolor{darkgray}\textcolor{white}{BVH-8} \\
    \multirow{4}{*}{\includegraphics[width=0.138\textwidth]{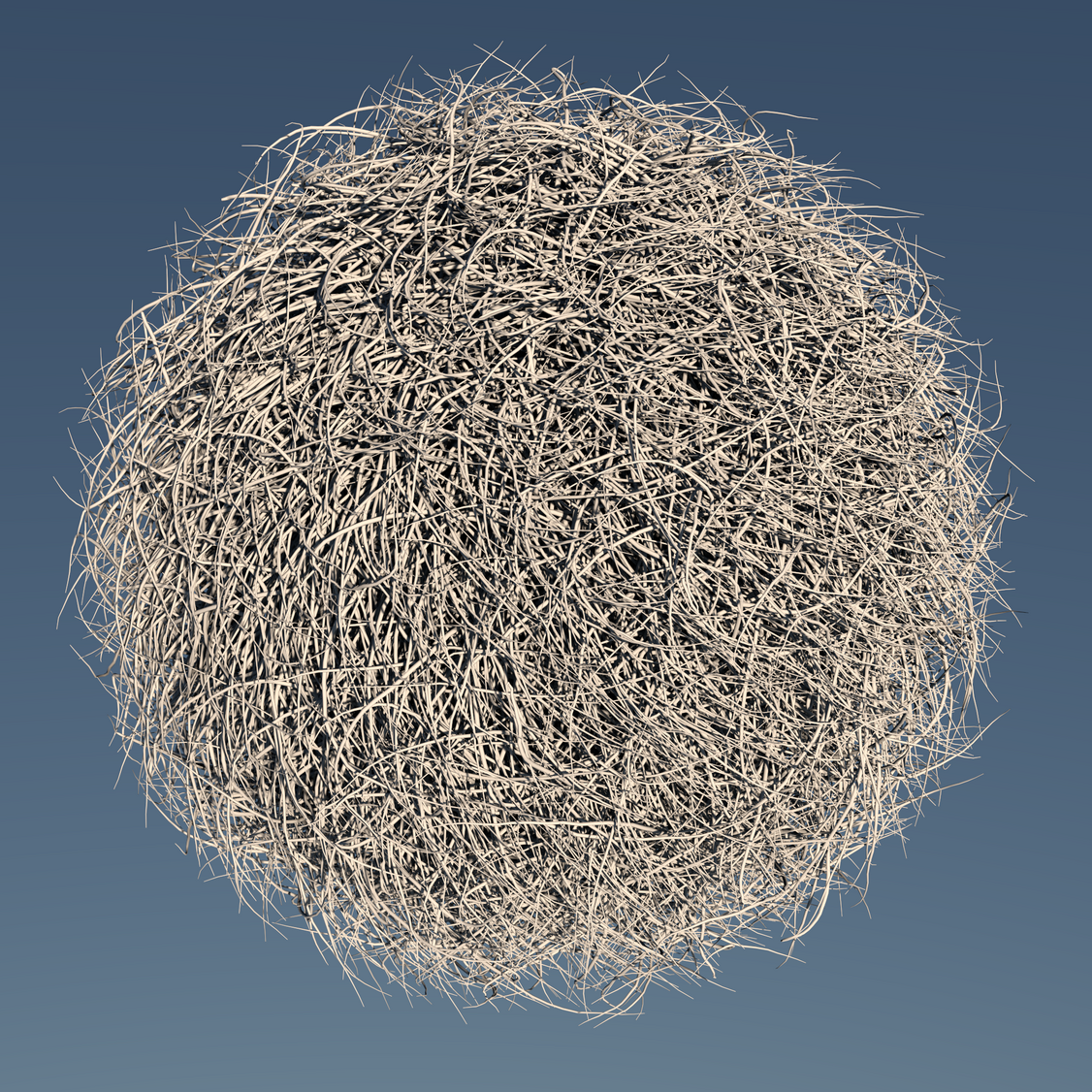}} & & \tiny AABB (Baseline) & 254.6 & 216.7 & 195.6 & 403 & 340 & 304 & 49.3 & 42.0 & 37.7\\
    & & {\tiny AABB $\rightarrow$ OBB$_\text{PC}$} & 141.5 & 115.3 & 101.7 & 226  & 187  & 158  & 34.0  & 28.8  & 25.8  \\
    & & \tiny (Roofline Per-Child) & \textcolor{teal}{\tiny$0.56\times$} & \textcolor{teal}{\tiny$0.53\times$} & \textcolor{teal}{\tiny$0.52\times$} & \textcolor{teal}{\tiny$0.59\times$} & \textcolor{teal}{\tiny$0.63\times$} & \textcolor{teal}{\tiny$0.64\times$} & \textcolor{teal}{\tiny$0.73\times$} & \textcolor{teal}{\tiny$0.73\times$} & \textcolor{teal}{\tiny$0.74\times$} \\
    & Hairball & {\tiny AABB $\rightarrow$ OBB$_\text{soCPU}$} & 153.5 & 128.3 & 116.0 & 236  & 213  & 196  & 35.9  & 30.6  & 27.7  \\
    &  \tiny[2.9M] & \tiny \textbf{Ours (Brute-Force)} & \textcolor{teal}{\tiny$0.60\times$} & \textcolor{teal}{\tiny$0.59\times$} & \textcolor{teal}{\tiny$0.59\times$} & \textcolor{teal}{\tiny$0.59\times$} & \textcolor{teal}{\tiny$0.63\times$} & \textcolor{teal}{\tiny$0.64\times$} & \textcolor{teal}{\tiny$0.73\times$} & \textcolor{teal}{\tiny$0.73\times$} & \textcolor{teal}{\tiny$0.74\times$} \\
    & & {\tiny {\tiny AABB $\rightarrow$ OBB$_\text{soGPU}$}} & 200.7 & 171.2 & 146.7 & 314  & 265  & 215  & 41.8  & 36.0  & 31.4  \\
    & & \tiny \textbf{Ours (GPU-based) }& \textcolor{teal}{\tiny$0.79\times$} & \textcolor{teal}{\tiny$0.79\times$} & \textcolor{teal}{\tiny$0.75\times$} & \textcolor{teal}{\tiny$0.78\times$} & \textcolor{teal}{\tiny$0.78\times$} & \textcolor{teal}{\tiny$0.71\times$} & \textcolor{teal}{\tiny$0.85\times$} & \textcolor{teal}{\tiny$0.86\times$} & \textcolor{teal}{\tiny$0.83\times$} \\
    \cline{2-12}
    \multirow{4}{*}{\includegraphics[width=0.138\textwidth]{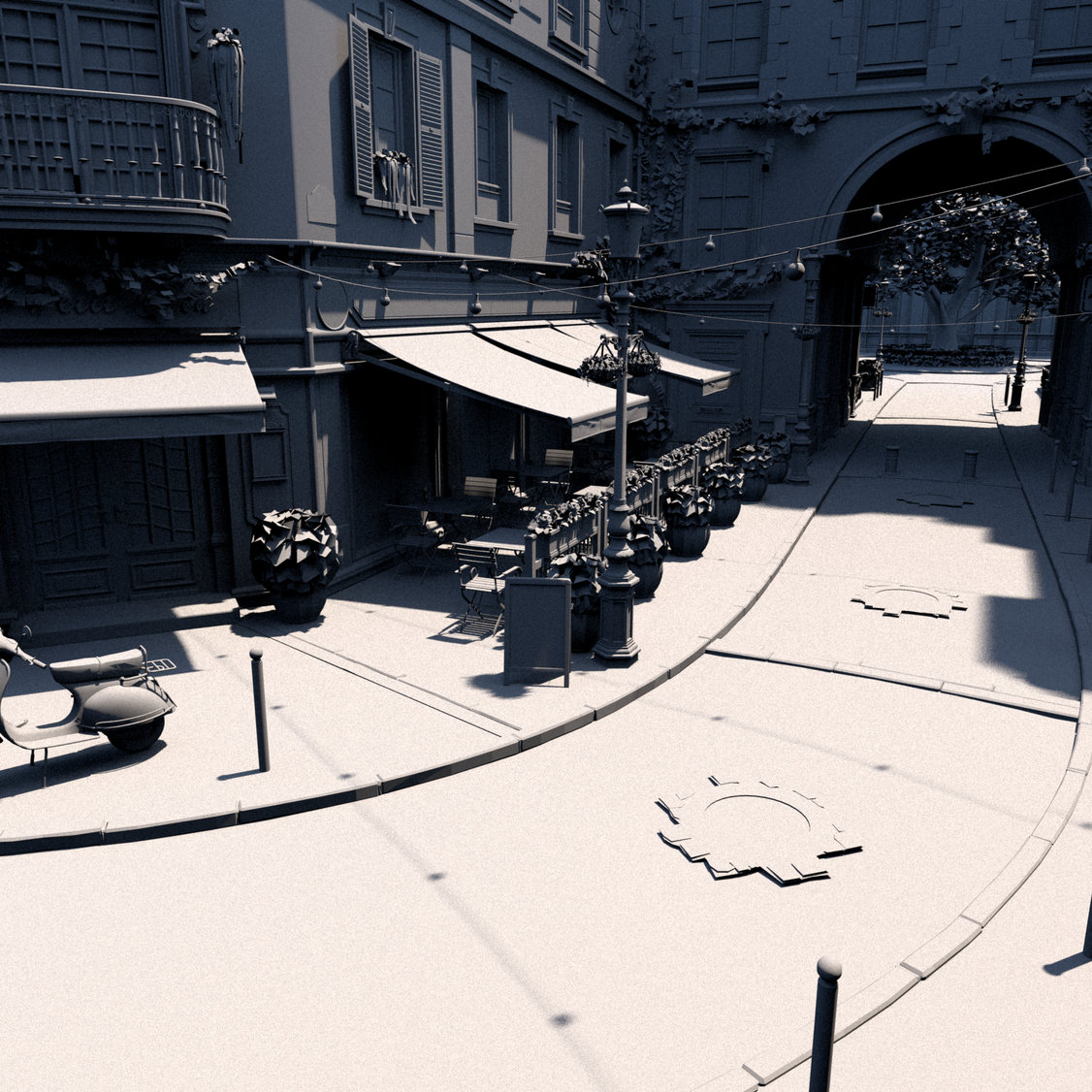}} & & \tiny AABB (Baseline) & 48.5 & 42.5 & 39.7 & 237 & 200 & 175 & 43.0 & 35.8 & 31.8 \\
    & & {\tiny AABB $\rightarrow$ OBB$_\text{PC}$} & 29.9 & 22.3 & 21.3 & 194  & 195  & 184  & 34.7  & 29.5  & 26.1  \\
    & Bistro & {\tiny \textcolor{teal}{Rel. Factor to AABB}} & \tiny\textcolor{teal}{$0.62\times$} & \tiny\textcolor{teal}{$0.52\times$} & \tiny\textcolor{teal}{$0.54\times$} & \tiny\textcolor{teal}{$0.82\times$} & \tiny\textcolor{teal}{$0.98\times$} & \tiny\textcolor{teal}{$1.05\times$} & \tiny\textcolor{teal}{$0.81\times$} & \tiny\textcolor{teal}{$0.82\times$} & \tiny\textcolor{teal}{$0.82\times$} \\
    & Exterior & {\tiny AABB $\rightarrow$ OBB$_\text{soCPU}$} & 26.7 & 22.7 & 20.8 & 207  & 183  & 151  & 38.1  & 31.7  & 28.1  \\
    & \tiny[2.8M] & & \tiny\textcolor{teal}{$0.55\times$} & \tiny\textcolor{teal}{$0.53\times$} & \tiny\textcolor{teal}{$0.52\times$} & \tiny\textcolor{teal}{$0.87\times$} & \tiny\textcolor{teal}{$0.92\times$} & \tiny\textcolor{teal}{$0.86\times$} & \tiny\textcolor{teal}{$0.89\times$} & \tiny\textcolor{teal}{$0.89\times$} & \tiny\textcolor{teal}{$0.88\times$} \\
    & & {\tiny AABB $\rightarrow$ OBB$_\text{soGPU}$} & 38.7 & 32.9	& 29.6 & 230  & 199  & 179  & 40.7  & 34.5  & 30.8  \\
    & & & \tiny\textcolor{teal}{$0.80\times$} & \tiny\textcolor{teal}{$0.77\times$} & \tiny\textcolor{teal}{$0.75\times$} & \tiny\textcolor{teal}{$0.97\times$} & \tiny\textcolor{teal}{$1.00\times$} & \tiny\textcolor{teal}{$1.02\times$} & \tiny\textcolor{teal}{$0.95\times$} & \tiny\textcolor{teal}{$0.96\times$} & \tiny\textcolor{teal}{$0.97\times$} \\
    \cline{2-12}
    \cline{2-12}
    \multirow{4}{*}{\includegraphics[width=0.138\textwidth]{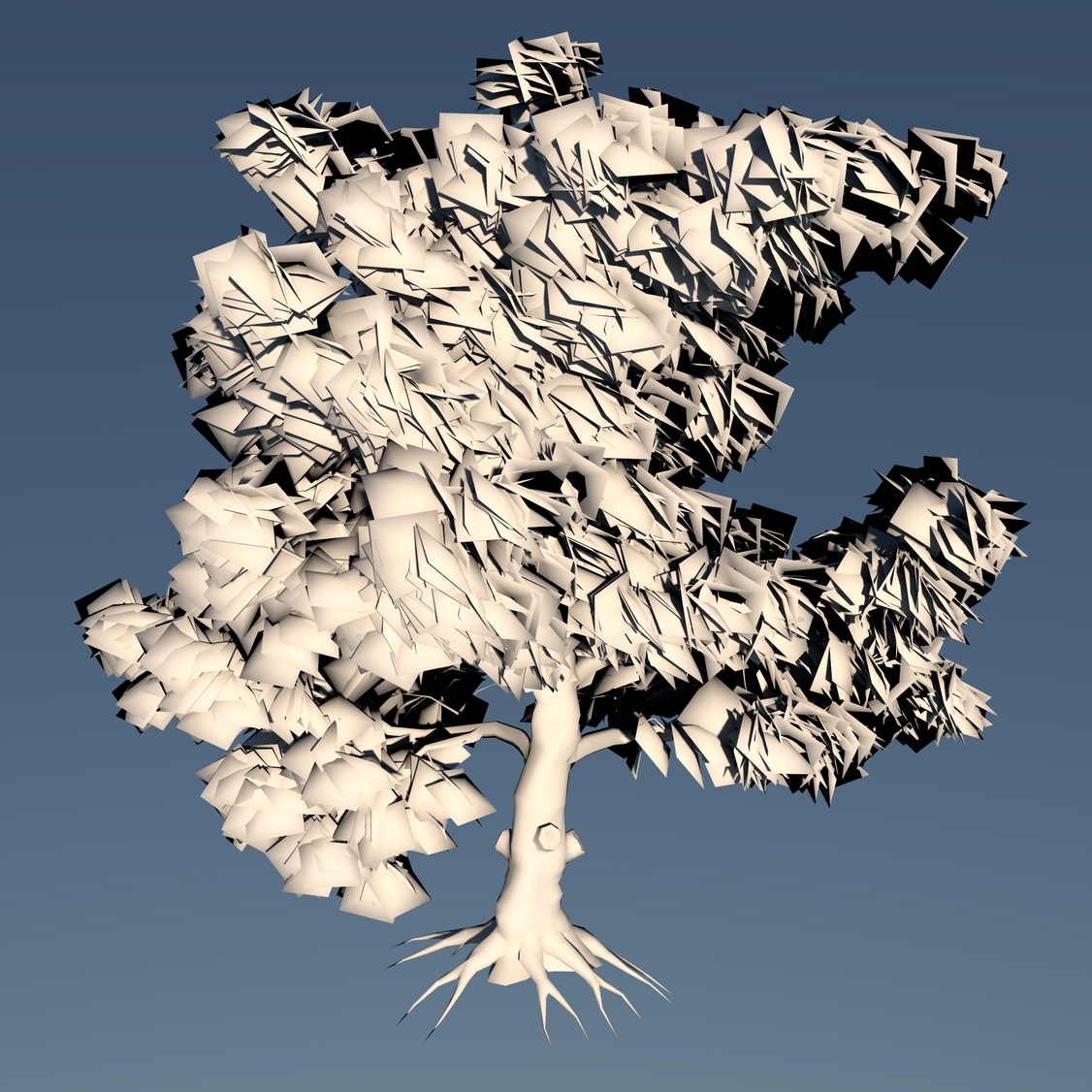}} & & \tiny AABB (Baseline) & 67.9 & 58.3 & 53.8 & 156 & 137 & 119 & 20.0 & 17.5 & 16.3 \\
    & & {\tiny AABB $\rightarrow$ OBB$_\text{PC}$} & 57.0 & 48.4 & 46.7 & 117  & 112  & 94  & 17.6  & 15.6  & 14.4  \\
    & & & \tiny\textcolor{teal}{$0.84\times$} & \tiny\textcolor{teal}{$0.83\times$} & \tiny\textcolor{teal}{$0.87\times$} & \tiny\textcolor{teal}{$0.75\times$} & \tiny\textcolor{teal}{$0.82\times$} & \tiny\textcolor{teal}{$0.79\times$} & \tiny\textcolor{teal}{$0.88\times$} & \tiny\textcolor{teal}{$0.89\times$} & \tiny\textcolor{teal}{$0.88\times$} \\
    & White Oak & {\tiny AABB $\rightarrow$ OBB$_\text{soCPU}$} & 61.9 &53.4 & 49.5 & 134  & 118  & 110  & 18.9  & 16.8  & 15.7  \\
    & \tiny [36K] & & \tiny\textcolor{teal}{$0.91\times$} & \tiny\textcolor{teal}{$0.92\times$} & \tiny\textcolor{teal}{$0.92\times$} & \tiny\textcolor{teal}{$0.86\times$} & \tiny\textcolor{teal}{$0.86\times$} & \tiny\textcolor{teal}{$0.92\times$} & \tiny\textcolor{teal}{$0.95\times$} & \tiny\textcolor{teal}{$0.96\times$} & \tiny\textcolor{teal}{$0.97\times$} \\
    & & {\tiny AABB $\rightarrow$ OBB$_\text{soGPU}$} & 66.1 & 56.8 & 52.6 & 150  & 132  & 113  & 19.7  & 17.3  & 16.1  \\
    & & & \tiny\textcolor{teal}{$0.97\times$} & \tiny\textcolor{teal}{$0.98\times$} & \tiny\textcolor{teal}{$0.98\times$}     & \tiny\textcolor{teal}{$0.96\times$} & \tiny\textcolor{teal}{$0.96\times$} & \tiny\textcolor{teal}{$0.95\times$} & \tiny\textcolor{teal}{$0.98\times$} & \tiny\textcolor{teal}{$0.99\times$} & \tiny\textcolor{teal}{$0.99\times$} \\
    \cline{2-12}
    \cline{2-12}
    \multirow{4}{*}{\includegraphics[width=0.138\textwidth]{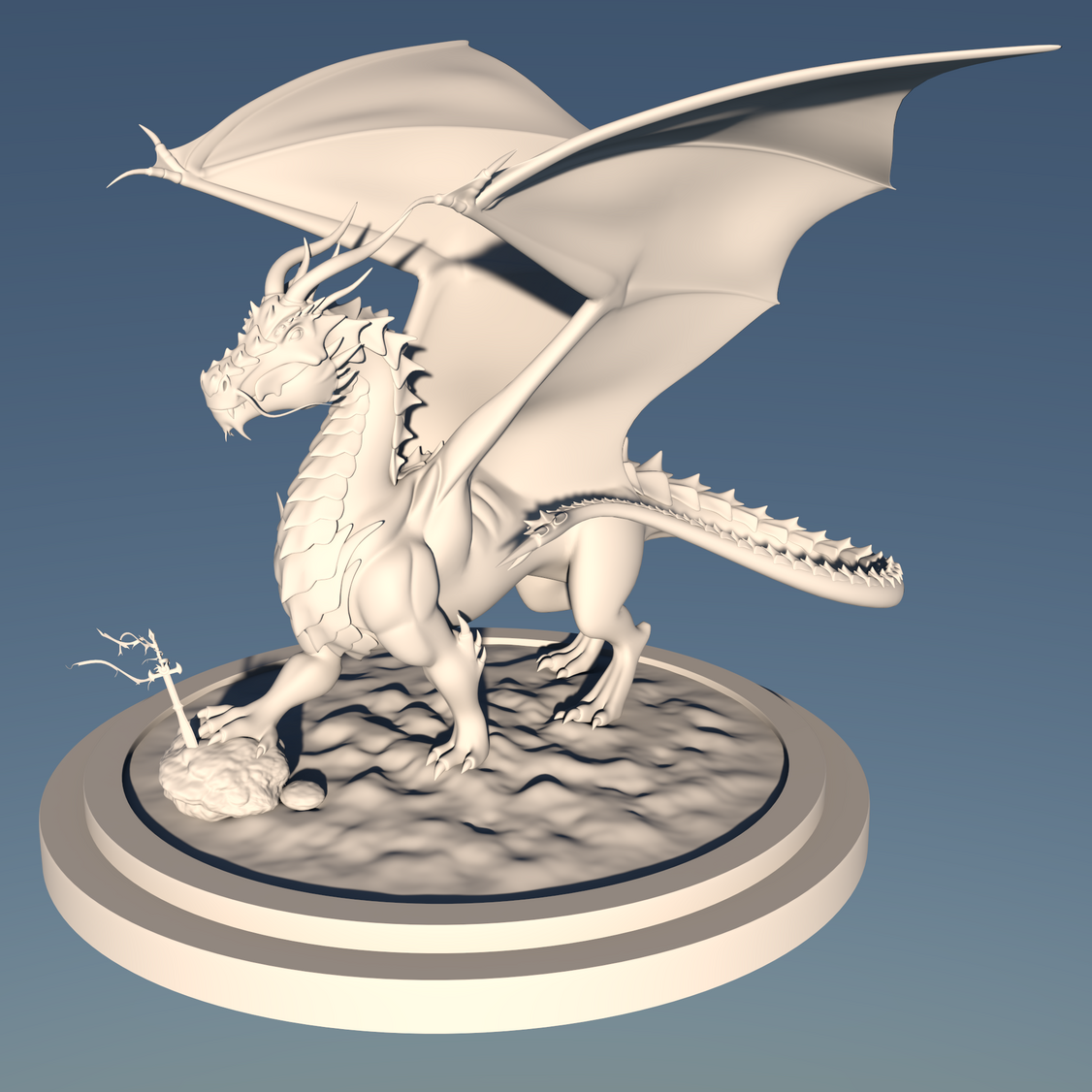}} & & \tiny AABB (Baseline) & 25.9 & 21.7 & 19.7 & 208 & 176 & 153 & 12.3 & 10.5 & 9.8 \\
    & & {\tiny AABB $\rightarrow$ OBB$_\text{PC}$} & 21.4 & 17.7 & 13.0 & 90  & 75  & 68  & 10.9  & 9.4  & 8.6  \\
    & & & \tiny\textcolor{teal}{$0.82\times$} & \tiny\textcolor{teal}{$0.81\times$} & \tiny\textcolor{teal}{$0.66\times$} & \tiny\textcolor{teal}{$0.43\times$} & \tiny\textcolor{teal}{$0.43\times$} & \tiny\textcolor{teal}{$0.44\times$} & \tiny\textcolor{teal}{$0.89\times$} & \tiny\textcolor{teal}{$0.89\times$} & \tiny\textcolor{teal}{$0.88\times$} \\
    & Dragon & {\tiny AABB $\rightarrow$ OBB$_\text{soCPU}$} & 22.4 & 18.8 & 17.0 & 94  & 81  & 71  & 11.3  & 9.8  & 9.0  \\
    & \tiny [830K] & & \tiny\textcolor{teal}{$0.86\times$} & \tiny\textcolor{teal}{$0.86\times$} & \tiny\textcolor{teal}{$0.86\times$} & \tiny\textcolor{teal}{$0.45\times$} & \tiny\textcolor{teal}{$0.46\times$} & \tiny\textcolor{teal}{$0.46\times$} & \tiny\textcolor{teal}{$0.92\times$} & \tiny\textcolor{teal}{$0.93\times$} & \tiny\textcolor{teal}{$0.92\times$} \\
    & & {\tiny AABB $\rightarrow$ OBB$_\text{soGPU}$} & 24.6 & 20.6 & 18.6 & 161  & 124  & 107  & 12.0  & 10.2  & 9.4  \\
    & & & \tiny\textcolor{teal}{$0.95\times$} & \tiny\textcolor{teal}{$0.95\times$} & \tiny\textcolor{teal}{$0.95\times$} & \tiny\textcolor{teal}{$0.77\times$} & \tiny\textcolor{teal}{$0.70\times$} & \tiny\textcolor{teal}{$0.70\times$} & \tiny\textcolor{teal}{$0.97\times$} & \tiny\textcolor{teal}{$0.97\times$} & \tiny\textcolor{teal}{$0.97\times$} \\
    \cline{2-12}
    \cline{2-12}
    \multirow{4}{*}{\includegraphics[width=0.138\textwidth]{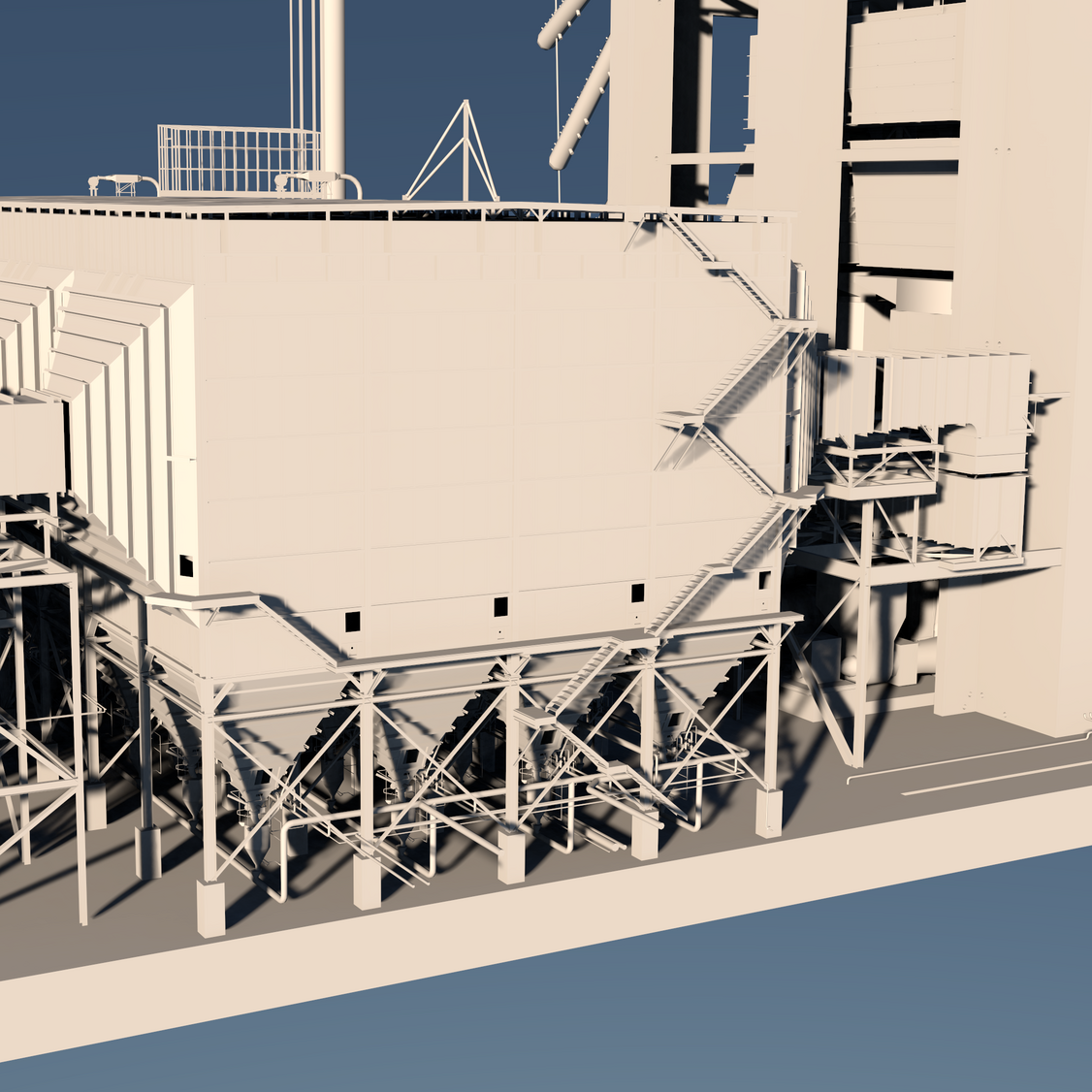}} & & \tiny AABB (Baseline) & 20.1 & 17.5 & 16.3 & 1788 & 1606 & 1574 & 48.1 & 38.5 & 36.6 \\
    & & {\tiny AABB $\rightarrow$ OBB$_\text{PC}$} & 14.3 & 12.2 & 11.2 & 369  & 326  & 306  & 36.2  & 28.3  & 26.6  \\
    & & & \tiny\textcolor{teal}{$0.71\times$} & \tiny\textcolor{teal}{$0.69\times$} & \tiny\textcolor{teal}{$0.69\times$} & \tiny\textcolor{teal}{$0.21\times$} & \tiny\textcolor{teal}{$0.20\times$} & \tiny\textcolor{teal}{$0.19\times$} & \tiny\textcolor{teal}{$0.75\times$} & \tiny\textcolor{teal}{$0.73\times$} & \tiny\textcolor{teal}{$0.73\times$} \\
    & Powerplant & {\tiny AABB $\rightarrow$ OBB$_\text{soCPU}$} & 14.6	& 12.5 & 11.5 & 373  & 333  & 315  & 38.1  & 29.4  & 27.5  \\
    & \tiny [12.8M]& & \tiny\textcolor{teal}{$0.72\times$} & \tiny\textcolor{teal}{$0.71\times$} & \tiny\textcolor{teal}{$0.71\times$} & \tiny\textcolor{teal}{$0.21\times$} & \tiny\textcolor{teal}{$0.21\times$} & \tiny\textcolor{teal}{$0.20\times$} & \tiny\textcolor{teal}{$0.79\times$} & \tiny\textcolor{teal}{$0.76\times$} & \tiny\textcolor{teal}{$0.75\times$} \\
    & & {\tiny AABB $\rightarrow$ OBB$_\text{soGPU}$} & 17.1 &14.5 & 13.5 & 831  & 725  & 565  & 39.5  & 30.6  & 28.5  \\
    & & & \tiny\textcolor{teal}{$0.85\times$} & \tiny\textcolor{teal}{$0.83\times$} & \tiny\textcolor{teal}{$0.83\times$} & \tiny\textcolor{teal}{$0.46\times$} & \tiny\textcolor{teal}{$0.45\times$} & \tiny\textcolor{teal}{$0.36\times$} & \tiny\textcolor{teal}{$0.82\times$} & \tiny\textcolor{teal}{$0.79\times$} & \tiny\textcolor{teal}{$0.78\times$} \\
    \cline{2-12}
    \cline{2-12}
  \end{tabular}
  \caption{BVH quality and traversal statistics for primary rays across different BVH widths for all datasets and OBB conversion algorithms.}
  \label{tab:data_sah_iters}
\end{table*}

{
\setlength{\tabcolsep}{0.5pt}
\begin{table*}[!h]
\centering
\begin{tabular}{cccl}
\cellcolor{darkgray}\textcolor{white}{AABB BVH-8} & \cellcolor{darkgray}\textcolor{white}{$\rightarrow$ OBB$_\text{soCPU}$ (Brute-Force)} &  \cellcolor{darkgray}\textcolor{white}{$\rightarrow$ OBB$_\text{soGPU}$ (Our Approach)} & \\
\includegraphics[width=.28\textwidth]{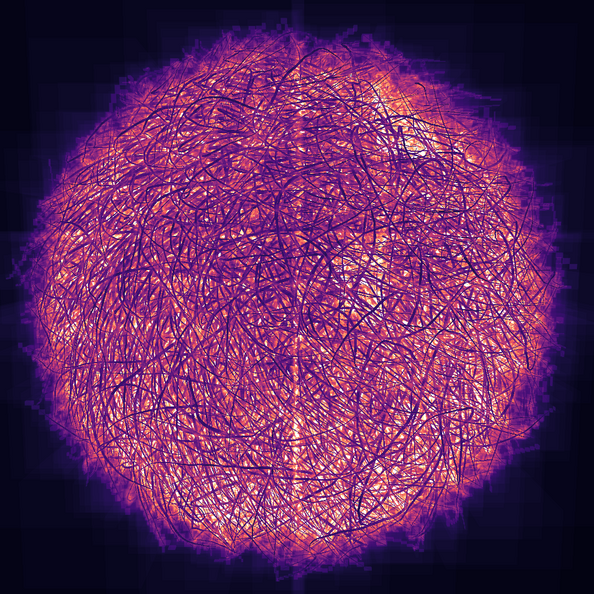} &
\includegraphics[width=.28\textwidth]{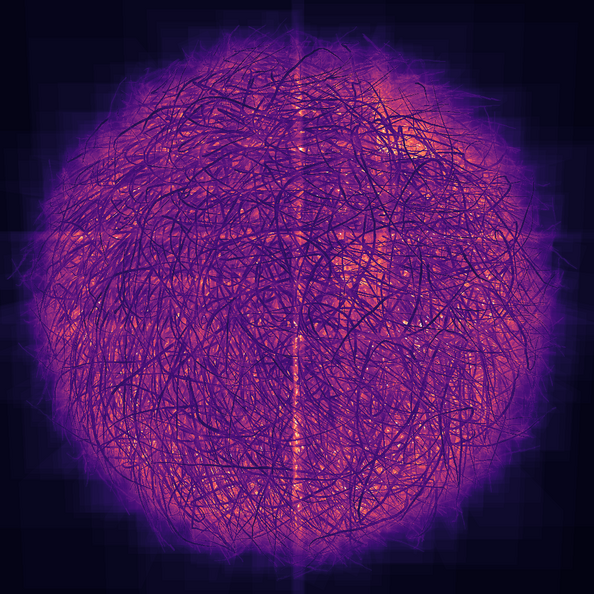} &
\includegraphics[width=.28\textwidth]{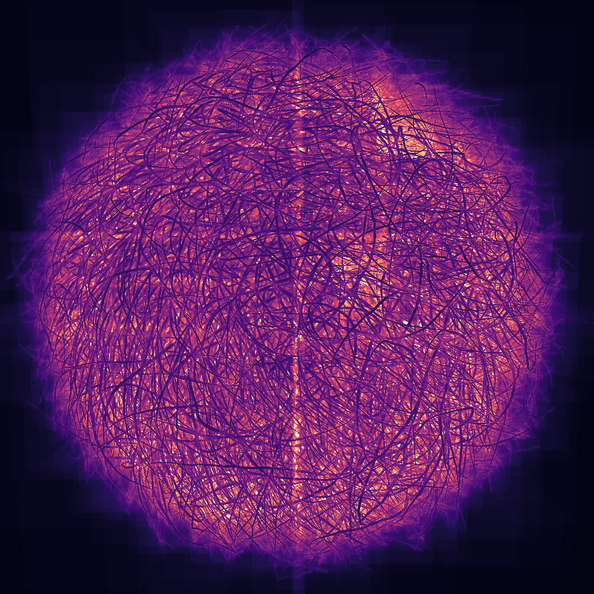} & \includegraphics[height=0.2\textheight]{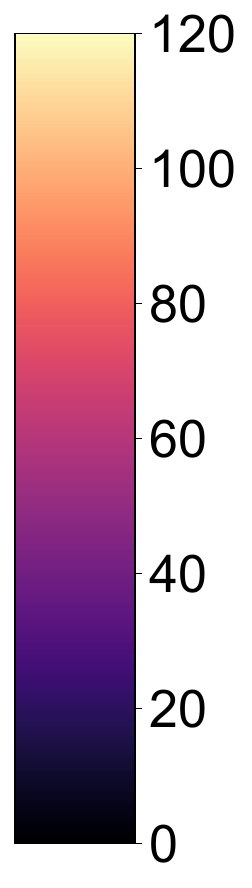} \\ [-2pt]
\includegraphics[width=.28\textwidth]{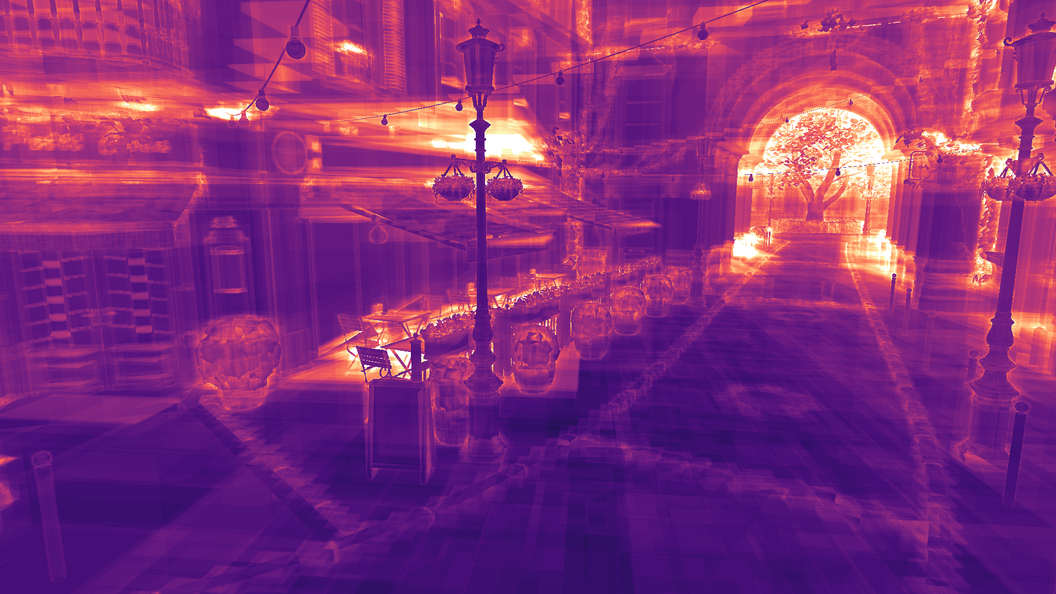} &
\includegraphics[width=.28\textwidth]{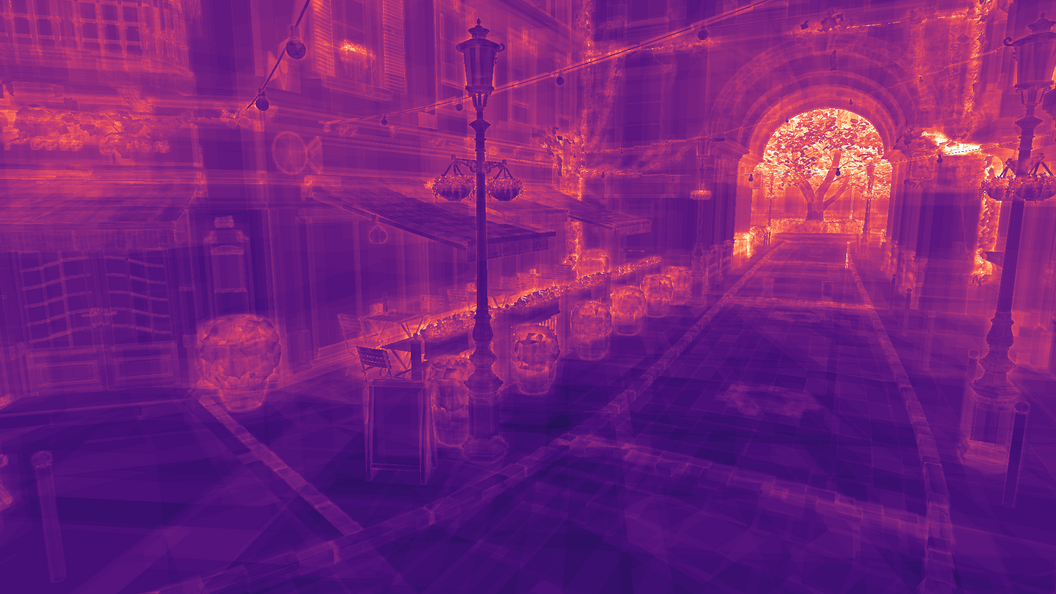} &
\includegraphics[width=.28\textwidth]{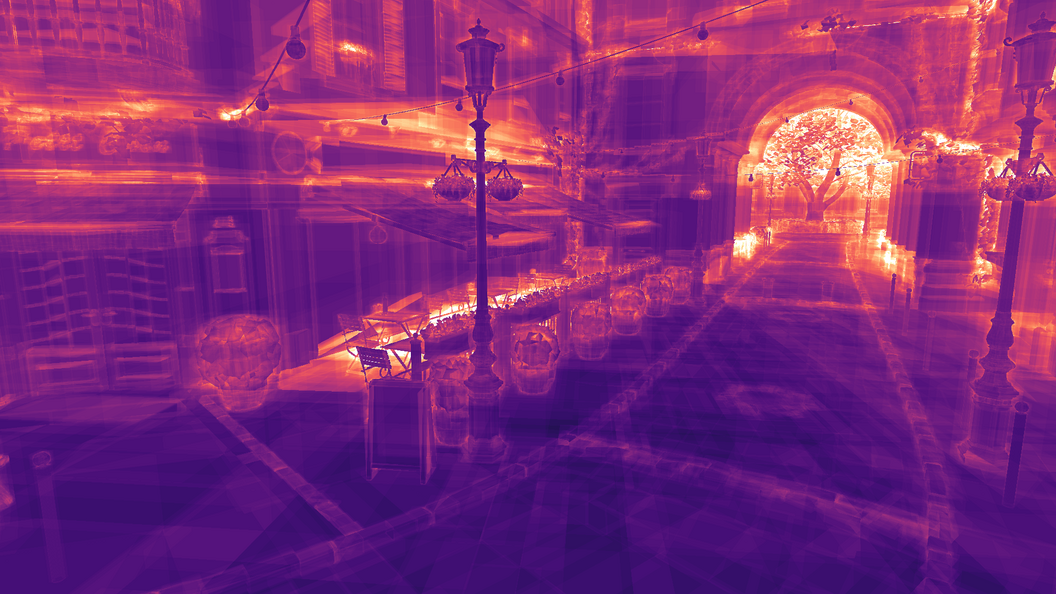} & \includegraphics[height=0.12\textheight]{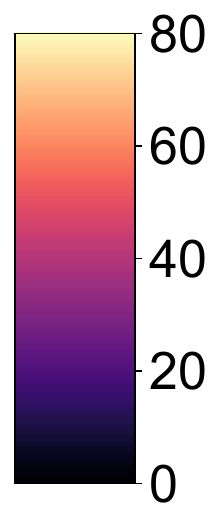} \\ [-2pt]
\includegraphics[width=.28\textwidth]{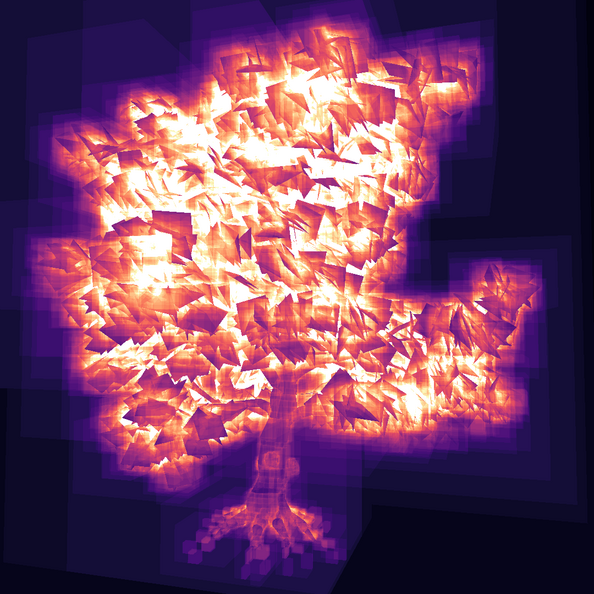} &
\includegraphics[width=.28\textwidth]{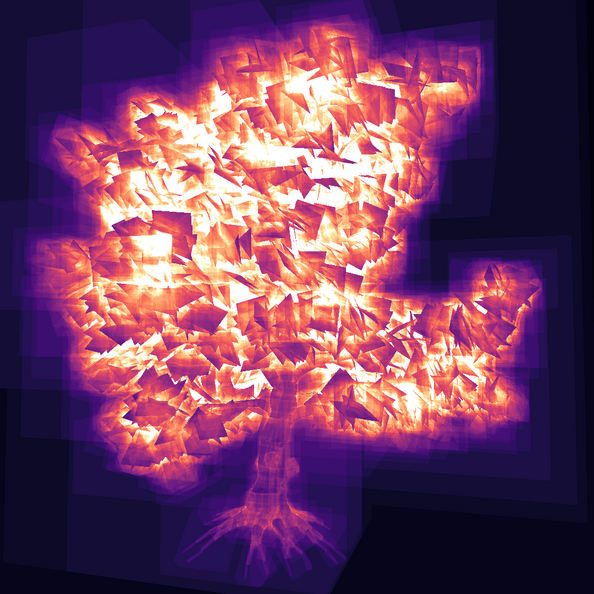} &
\includegraphics[width=.28\textwidth]{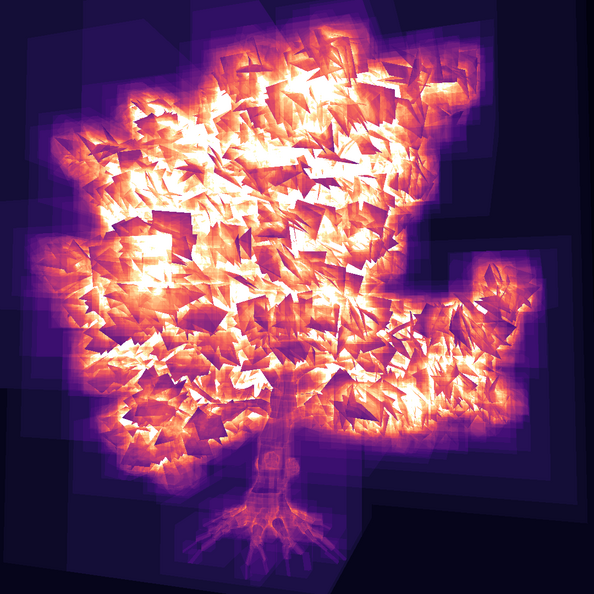}  & \includegraphics[height=0.2\textheight]{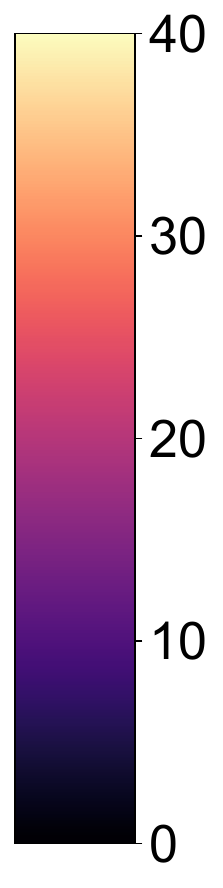} \\ [-2pt]
\includegraphics[width=.28\textwidth]{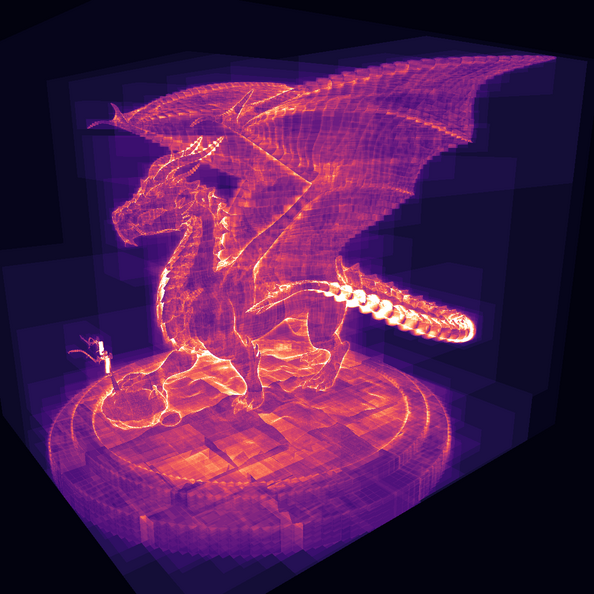} &
\includegraphics[width=.28\textwidth]{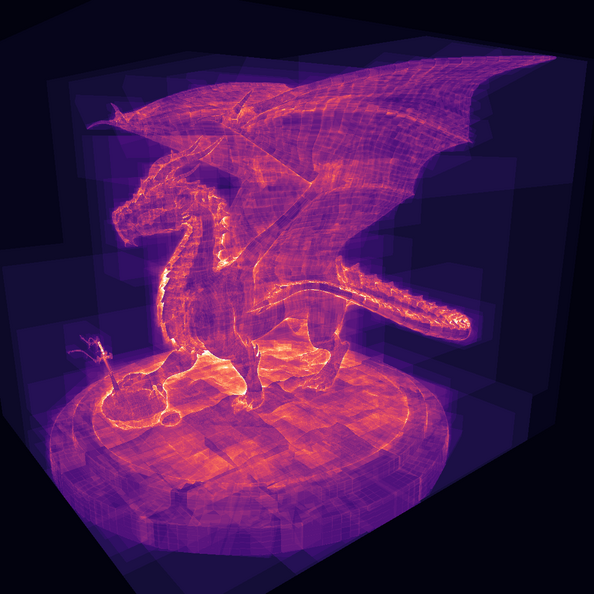} &
\includegraphics[width=.28\textwidth]{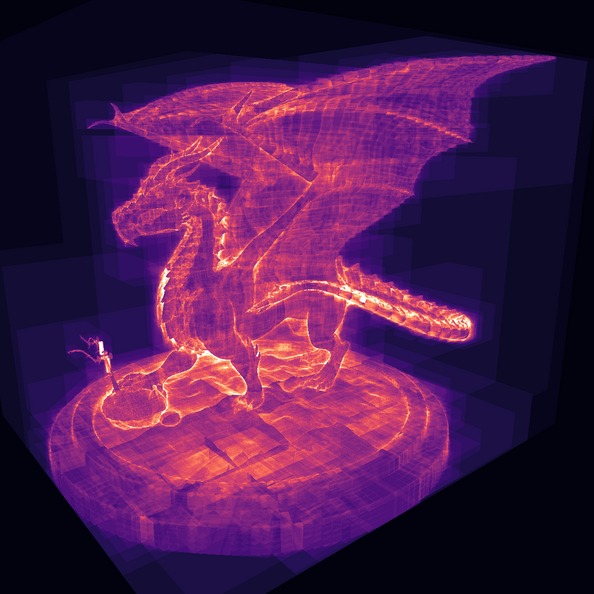}  & \includegraphics[height=0.2\textheight]{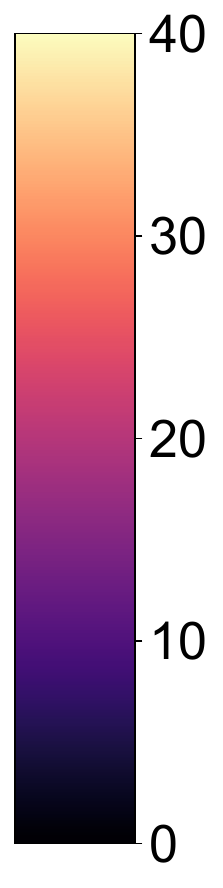} \\ [-2pt]
\includegraphics[width=.28\textwidth]{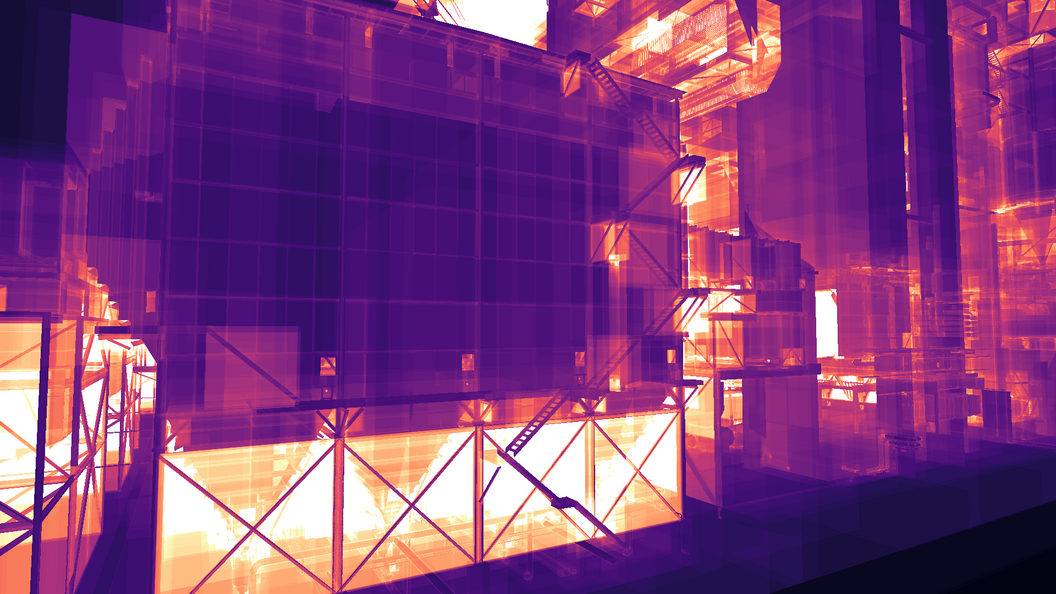} &
\includegraphics[width=.28\textwidth]{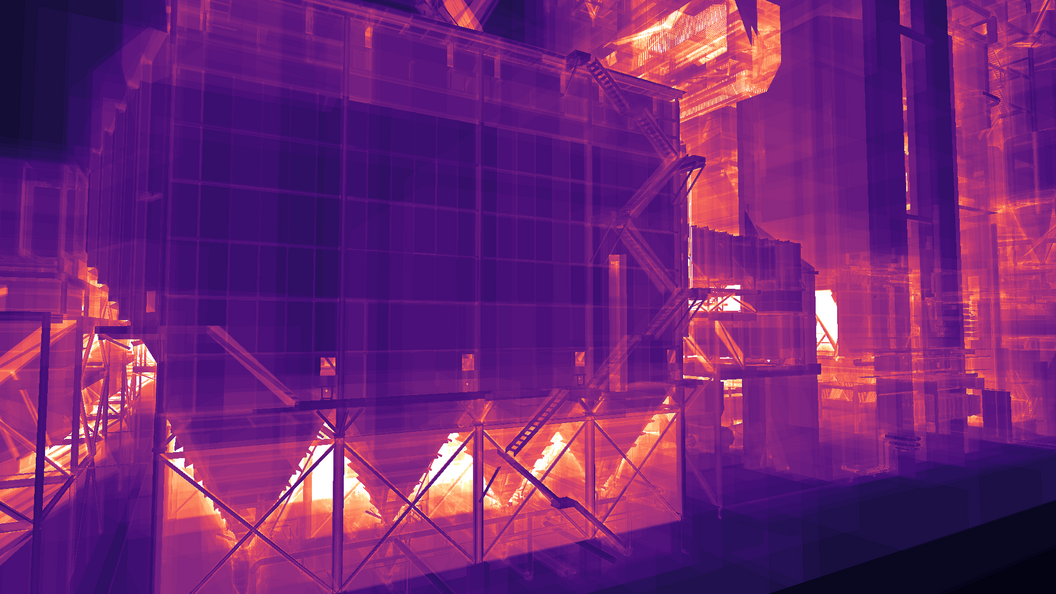} &
\includegraphics[width=.28\textwidth]{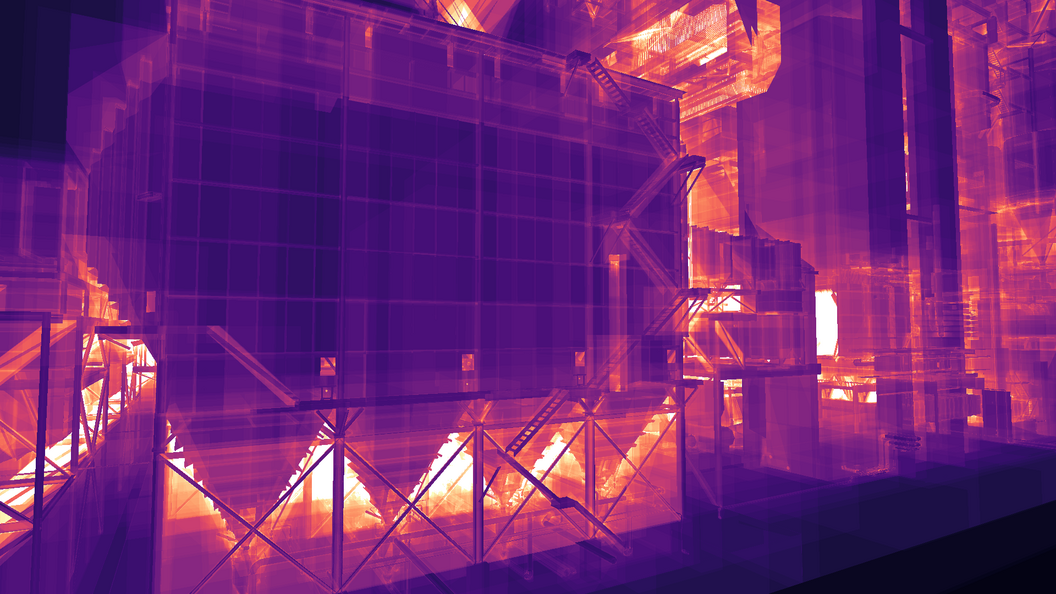} & \includegraphics[height=0.12\textheight]{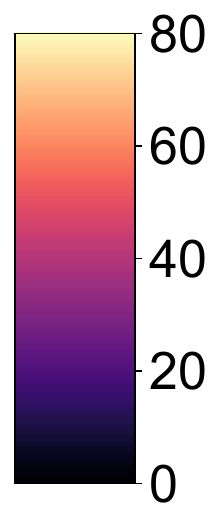} \\
\end{tabular}
\caption{Visualization of traversal loop iteration counts at each pixel for primary rays for our CPU-based and GPU-based DOBB BVHs compared to the baseline AABB BVH.}
\label{tab:bvh8_iters}
\end{table*}
}

\vspace{-8pt}
\subsection{Memory Consumption}
We store \emph{k} floats for each plane of a given \emph{k}-DOP, defaulting to 26 floats for a total of 104 bytes. Our worst-case allocation is $(N-1) \cdot 104{b}$ for $N - 1$ interior nodes. 
We also maintained a list of sorted node indices from which to start the refit kernel, which requires $N \cdot 4$ bytes where $N$ is the maximum number of internal nodes.
Finally, 3 lookup tables were used for our precomputed set of orientations.
To map our OBB matrix index to a $3\times3$ rotation, we relied on a 2-level lookup table that uses 960B total, and to map the rotation to an OBB matrix index, we use a 3D look up table that uses 4096B.

\vspace{-8pt}
\subsection{Build Time and Real-Time Performance}
We evaluated the build time of our OBB$_\text{soGPU}$ approach including all AABB build passes and measured final ray tracing performance by tracing both primary and secondary rays through each scene from 25 different viewpoints. Results for BVH-8 are presented in Table~\ref{tab:hw_results_bvh8} and highlighted in Fig.~\ref{fig:teaser}.
Overall, OBB$_\text{soGPU}$ introduces a moderate build time overhead of approximately 10--15\% relative to the baseline BVH construction pipeline. 
In absolute terms, this corresponds to an additional 0.1~ms for the smallest scene and up to 5~ms for the largest scene.
We consider this acceptable for real-time applications, particularly for static content.
In return, our DOBB-BVHs deliver notable run-time improvements. For primary rays, we observe an average speedup of 18.5\%, with gains reaching up to 43\% for primary rays. 
The most substantial improvements appear for secondary rays, particularly in global illumination (GI) workloads where rays originate from geometry surfaces and exhibit high divergence. 
In these cases, performance increases by up to 65\%.
In comparison, OBB$_\text{soCPU}$ achieves promising traversal speedups of up to 76\% for primary rays and 106\% for GI rays.
These results reinforce our earlier findings: scenes with predominantly non-axis-aligned geometry benefit most, due to significant reductions in maximum traversal iterations. 
As supported by both SAH and iteration count analyses, these performance gains hold consistently across all tested BVH branching factors and ray types.
A breakdown of BVH construction phases is provided in Fig.~\ref{fig:obb-build-time}, while Table~\ref{tab:hw_results} in App.~\ref{app:breakdown} contains details build times and rendering performance across all branching factors.
\vspace{-8pt}
\section{Conclusion}
\label{sec:conclusion}
In this paper, we have introduced DOBB-BVH, a novel bounding volume hierarchy based on OBBs, which maintains the same rotation between all wide hierarchy node children through a fixed set of discrete rotations. 
Our approach employs an efficient GPU-based post-processing algorithm to identify and compute OBBs from existing AABB BVHs, leveraging hardware-accelerated ray-OBB intersections.
To evaluate the effectiveness of DOBB-BVH, we have conducted a detailed performance study demonstrating significant improvements in ray traversal: up to an 80\% reduction in worst-case traversal steps and a 10–20\% decrease in average iterations per ray. 
Despite extra a build-time overhead of 10–12\%, our method achieves real-time ray tracing performance improvements of 22.7\% on average, with gains up to 43\% for primary rays and 65\% for secondary rays. These results are enabled by an efficient mapping of geometry to optimal OBB candidates and an improved refitting strategy using apex point maps as proxies.

In the future, our primary objective is to close the noticeable performance gap of $\sim$20\% in traversal efficiency and $\sim$40\% in overall performance compared to a brute-force CPU baseline.
Additional promising research directions include developing improved cost heuristics, optimizing the mapping of geometry and child nodes to OBB candidates, and refining the discrete rotation set.
We additionally plan to combine the implementation of DOBB with triangle splitting.
We also see potential in hybrid bounding volume approaches, such as incorporating AABBs and OBBs within a single node (e.g., 12-DOPs), to better adapt to heterogeneous scene content.
Modifying early stages of the BVH build process can yield benefits. For example, constructing OBB-aware quads or orientation-coherent clusters could improve fitting quality.
Finally, we aim to investigate efficient OBB construction techniques tailored for dynamic or updatable BVHs, enabling broader applicability in real-time and interactive rendering scenarios.

\paragraph*{Acknowledgements}
We thank our colleague Julien Beasley for his work on look-up tables. The \emph{Dragon} model was obtained from Benedikt Bitterli’s Rendering Resources site, the \emph{Bistro Exterior} model is courtesy of Amazon Lumberyard. All other scenes were obtained from Morgan McGuire's resource archive. 
Datasets and bounding box images were rendered using the AMD Capsaicin Framework~\cite{capsaicin2023} and Mitsuba~\cite{mitsuba2010}.

\vspace*{-8pt}
\subsection*{Copyright Notice and Trademarks}
\textcopyright\space 2025 Advanced Micro Devices, Inc. All rights reserved. AMD, Ryzen, Radeon, and combinations thereof are trademarks of Advanced Micro Devices, Inc. Other product names used in this publication are for identification purposes only and may be trademarks of their respective companies.

\vspace{-8pt}
\bibliographystyle{eg-alpha-doi} 
\bibliography{main}
\newpage
\appendix 
\onecolumn
\section{Breakdown of BVH Build Phases}
\label{app:breakdown}
\begin{figure}[H]
	\onecolumn
    \centering
    \includegraphics[width=1\linewidth]{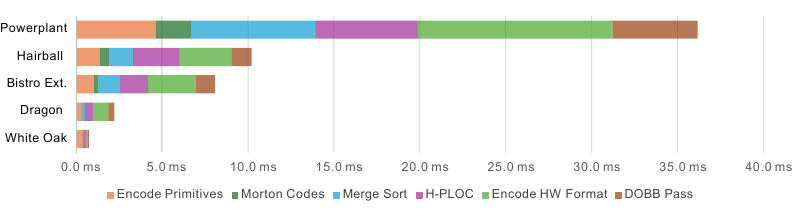}
    \caption{Breakdown of BVH-8 build phases across our set of test scenes, showcasing the impact of DOBB-BVH.}
    \label{fig:obb-build-time}
\end{figure}

We measured the execution times of all individual steps in our BVH construction pipeline, as shown in Fig.~\ref{fig:obb-build-time}, and report the results in milliseconds in Table~\ref{tab:hw_build_results}. 
For scenes containing up to three million triangles, our DOBB post-processing step takes up to 1.15ms and increases to approximately 5ms for scenes with 13 million triangles.

\begin{table}[!hb]
    \centering
    \small
    \setlength{\tabcolsep}{8pt}
    {
    \begin{tabular}{|l|l|l|l|l|l|}
    \hline
        {\cellcolor{black}\textcolor{white}{Build Phases}} &{\cellcolor{black}\textcolor{white}{White Oak}} & {\cellcolor{black}\textcolor{white}{Dragon}} & {\cellcolor{black}\textcolor{white}{Bistro Ext.}} & {\cellcolor{black}\textcolor{white}{Hairball}} & {\cellcolor{black}\textcolor{white}{Powerplant}} \\ \hline
        Encode Primitives &  0.39 ms &  0.33 ms &  1.04 ms &  1.36 ms &  4.62 ms \\ \hline
        Morton Codes &  0.01 ms &  0.05 ms &  0.26 ms &  0.58 ms &  2.05 ms \\ \hline
        Merge Sort &  0.04 ms &  0.13 ms &  1.21 ms &  1.38 ms &  7.29 ms \\ \hline
        H-PLOC &  0.16 ms &  0.51 ms &  1.67 ms &  2.67 ms &  5.94 ms \\ \hline
        Encode HW Format &  0.09 ms &  0.86 ms &  2.77 ms &  3.07 ms &  11.32 ms \\ \hline
        DOBB Pass &  0.05 ms &  0.35 ms &  1.13 ms &  1.15 ms &  4.95 ms \\ \hline
    \end{tabular}
    }
    \caption{Results of BVH-8 build phases in Figure~\ref{fig:obb-build-time}.}
    \label{tab:hw_build_results}
\end{table}

\section{Analysis of Traversal Performance for Primary and Secondary Rays}
\label{app:analysis}
We further evaluated the build times and traversal performance of our DOBB algorithm for both primary and secondary rays. Table~\ref{tab:hw_results} presents the build and traversal performance across BVHs with varying branching factors. Detailed analysis of BVH quality and traversal behavior for secondary (global illumination) rays is provided in Table~\ref{tab:data_sah_iters_gi} and Table~\ref{tab:bvh8_iters_gi}.
\begin{table*}[!h]
    \centering
    \small
    \begin{tabular}{|l|l|l|l|l|l|l|l|l|l|l|}
        \hline 
        \cellcolor{black} & \cellcolor{black} & \multicolumn{3}{c|}{\cellcolor{black}\textcolor{white}{Build Time}} & \multicolumn{3}{c|}{\cellcolor{black}\textcolor{white}{Primary}} & \multicolumn{3}{c|}{\cellcolor{black}\textcolor{white}{Secondary}} \\
    \cellcolor{black} & \cellcolor{black} & \multicolumn{3}{c|}{\cellcolor{black}\textcolor{white}{ms}} & \multicolumn{3}{c|}{\cellcolor{black}\textcolor{white}{GRays/s}} & \multicolumn{3}{c|}{\cellcolor{black}\textcolor{white}{GRays/s}} \\
    \multirow{-3}{*}{\cellcolor{black}\textcolor{white}{Scene}} & \multirow{-3}{*}{\cellcolor{black}\textcolor{white}{\tiny Approach}} & \multicolumn{1}{c|}{\cellcolor{darkgray}\textcolor{white}{BVH-4}} & \multicolumn{1}{c|}{\cellcolor{darkgray}\textcolor{white}{BVH-6}} & \multicolumn{1}{c|}{\cellcolor{darkgray}\textcolor{white}{BVH-8}} & \multicolumn{1}{c|}{\cellcolor{darkgray}\textcolor{white}{BVH-4}} & \multicolumn{1}{c|}{\cellcolor{darkgray}\textcolor{white}{BVH-6}} & \multicolumn{1}{c|}{\cellcolor{darkgray}\textcolor{white}{BVH-8}} & \multicolumn{1}{c|}{\cellcolor{darkgray}\textcolor{white}{BVH-4}} & \multicolumn{1}{c|}{\cellcolor{darkgray}\textcolor{white}{BVH-6}} & \multicolumn{1}{c|}{\cellcolor{darkgray}\textcolor{white}{BVH-8}} \\
        ~ & \tiny AABB & 12.61 & 12.07 & 11.62 & 2.27 & 2.63 &  2.91 &  0.89 &  1.06 &  1.16 \\
        \multirow{-2}{*}{Hairball} & \tiny OBB$_\text{soGPU}$ & 14.96 \tiny\textcolor{purple}{(1.19$\times$)} & 13.26 \tiny\textcolor{purple}{(1.10$\times$)} & 12.82 \tiny\textcolor{purple}{(1.10$\times$)} & \textbf{3.19 \textcolor{teal}{\tiny(1.40$\times$)}} & \textbf{3.62 \textcolor{teal}{\tiny(1.38$\times$)}} & \textbf{4.16 \textcolor{teal}{\tiny(\textbf{1.43$\times$})}} & \textbf{1.43 \tiny\textcolor{teal}{(1.35$\times$)}} &  \textbf{1.66 \tiny\textcolor{teal}{(1.43$\times$)}} & \textbf{1.66} \tiny\textcolor{teal}{(\textbf{1.43$\times$})}\\ \hline
        ~ & \tiny AABB & 13.49 & 12.31 & 12.16 &  2.82 &  3.29 &  3.63 &  1.14 &  1.29 &  1.34 \\  
        \multirow{-2}{*}{Bistro Ext.} & \tiny OBB$_\text{soGPU}$ & 15.57 \tiny\textcolor{purple}{(1.26$\times$)} & 13.56 \tiny\textcolor{purple}{(1.10$\times$)} & 13.56 \tiny\textcolor{purple}{(1.12$\times$)} &  3.22 \tiny\textcolor{teal}{(1.14$\times$)} &  3.78 \tiny\textcolor{teal}{(1.15$\times$)} &  4.27 \tiny\textcolor{teal}{(1.18$\times$)} &  1.56 \tiny\textcolor{teal}{(1.36$\times$)} &  \textbf{1.82 \tiny\textcolor{teal}{(1.41$\times$)}} &  \textbf{1.97} \tiny\textcolor{teal}{(\textbf{1.47$\times$})} \\ \hline
        ~ & \tiny AABB & 0.70 & 0.55 & 0.65 &  6.57 &  7.49 &  7.95 &  2.39 &  2.70 &  2.82 \\ 
        \multirow{-2}{*}{White Oak} & \tiny OBB$_\text{soGPU}$ & 0.76 \tiny\textcolor{purple}{(1.09$\times$)} & 0.687 \tiny\textcolor{purple}{(1.25$\times$)} & 0.71 \tiny\textcolor{purple}{(1.09$\times$)} &  6.64 \tiny\textcolor{teal}{(1.01$\times$)} &  7.61 \tiny\textcolor{teal}{(1.02$\times$)} & 8.04 \tiny\textcolor{teal}{(1.01$\times$)} &  2.48 \tiny\textcolor{teal}{(1.04$\times$)} &  2.80 \tiny\textcolor{teal}{(1.03$\times$)} & 2.93 \tiny\textcolor{teal}{(1.04$\times$)} \\ \hline
        ~ & \tiny AABB & 2.57 & 2.49 & 2.57 &  17.78 &  19.99 &  20.93 &  2.87 &  3.17 &  3.26 \\ 
        \multirow{-2}{*} {Dragon} & \tiny OBB$_\text{soGPU}$ & 3.22 \tiny\textcolor{purple}{(1.25$\times$)} & 3.09 \tiny\textcolor{purple}{(1.24$\times$)} & 3.02x \tiny\textcolor{purple}{(1.18$\times$)} &  18.19 \tiny\textcolor{teal}{(1.02$\times$)} &  20.41 \tiny\textcolor{teal}{(1.02$\times$)} & 21.08 \tiny\textcolor{teal}{(1.01$\times$)} & 3.00 \tiny\textcolor{teal}{(1.04$\times$)} &  3.26 \tiny\textcolor{teal}{(1.03$\times$)} & 3.37 \tiny\textcolor{teal}{(1.03$\times$)} \\ \hline
        ~ & \tiny AABB & 51.02 & 49.50 & 48.46 & 3.54 &  3.92 &  4.09 &  1.73 &  2.11 &  1.87 \\ 
         \multirow{-2}{*}{Powerplant} & \tiny OBB$_\text{soGPU}$ & 60.33 \tiny\textcolor{purple}{(1.18$\times$)} & 55.37 \tiny\textcolor{purple}{(1.12$\times$)} & 55.20 \tiny\textcolor{purple}{(1.08$\times$)} &  \textbf{5.04 \tiny\textcolor{teal}{(1.42$\times$)}} &  \textbf{6.37 \tiny\textcolor{teal}{(1.63$\times$)}} & \textbf{5.32} \tiny\textcolor{teal}{(\textbf{1.30$\times$})} &  2.29 \tiny\textcolor{teal}{(1.33$\times$)} &  \textbf{3.08 \tiny\textcolor{teal}{(1.46$\times$)}} & \textbf{3.09} \tiny\textcolor{teal}{(\textbf{1.65$\times$})} \\ \hline
    \end{tabular}
    \caption{Hardware results of our GPU-based DOBB BVHs build algorithm across all branching factors 4--8 compared against an AABB tree and the relative factor within brackets (\textcolor{purple}{increased build time} or \textcolor{teal}{improved performance}).}
    \label{tab:hw_results}
\end{table*}

\begin{table*}[!h]
  \centering
  \small
  \begin{tabular}{c|c|c||c|c|c||c|c|c|}
    \cline{2-9}
    & \cellcolor{black}\textcolor{white}{Scene} & \cellcolor{black}\textcolor{white}{Approach} & \multicolumn{3}{c||}{\cellcolor{black}\textcolor{white}{Max. Iterations / Ray}} & \multicolumn{3}{c|}{\cellcolor{black}\textcolor{white}{Avg. Iterations / Ray}}\\
    & \cellcolor{darkgray}\textcolor{white}{[\#Triangles]} &  & \cellcolor{darkgray}\textcolor{white}{BVH-4} & \cellcolor{darkgray}\textcolor{white}{BVH-6} & \cellcolor{darkgray}\textcolor{white}{BVH-8} & \cellcolor{darkgray}\textcolor{white}{BVH-4} & \cellcolor{darkgray}\textcolor{white}{BVH-6} & \cellcolor{darkgray}\textcolor{white}{BVH-8} \\
    \multirow{4}{*}{\includegraphics[width=0.158\textwidth]{screenshot_hairball_mitsuba.png}} & & & & & & & &  \\
    & & \multirow{-2}{*}{\tiny AABB (Baseline)} & \multirow{-2}{*}{3735} & \multirow{-2}{*}{409} & \multirow{-2}{*}{378} & \multirow{-2}{*}{34.3} & \multirow{-2}{*}{28.9} & \multirow{-2}{*}{25.9} \\
    & & {\tiny AABB $\rightarrow$ OBB$_\text{PC}$} & 212  & 159  & 151  & 20.8  & 17.2  & 15.4  \\
    & Hairball & \tiny (Roofline Per-Child) & \tiny\textcolor{teal}{$0.06\times$} & \tiny\textcolor{teal}{$0.39\times$} & \tiny\textcolor{teal}{$0.40\times$} & \tiny\textcolor{teal}{$0.61\times$} & \tiny\textcolor{teal}{$0.60\times$} & \tiny\textcolor{teal}{$0.60\times$} \\
    & [2.9M] & {\tiny AABB $\rightarrow$ OBB$_\text{soCPU}$} & 346  & 180  & 176  & 22.2  & 18.7  & 16.9  \\
    & & \tiny \textbf{Ours (Brute-Force)} & \tiny\textcolor{teal}{$0.09\times$} & \tiny\textcolor{teal}{$0.44\times$} & \tiny\textcolor{teal}{$0.47\times$} & \tiny\textcolor{teal}{$0.65\times$} & \tiny\textcolor{teal}{$0.65\times$} & \tiny\textcolor{teal}{$0.65\times$} \\
    & & {\tiny {\tiny AABB $\rightarrow$ OBB$_\text{soGPU}$}} & 1050  & 268  & 247  & 27.4  & 23.4  & 20.1  \\
    & & \tiny \textbf{Ours (GPU-based) } & \tiny\textcolor{teal}{$0.28\times$} & \tiny\textcolor{teal}{$0.66\times$} & \tiny\textcolor{teal}{$0.65\times$} & \tiny\textcolor{teal}{$0.80\times$} & \tiny\textcolor{teal}{$0.81\times$} & \tiny\textcolor{teal}{$0.78\times$} \\
    \cline{2-9}
    \cline{2-9}
    \multirow{4}{*}{\includegraphics[width=0.158\textwidth]{screenshot_bistro_ext_mitsuba.png}} & & & & & & & &  \\
    & & \multirow{-2}{*}{\tiny AABB (Baseline)} & \multirow{-2}{*}{524} & \multirow{-2}{*}{492} & \multirow{-2}{*}{463} & \multirow{-2}{*}{61.4} & \multirow{-2}{*}{52.4} & \multirow{-2}{*}{48.3} \\ 
    & & {\tiny AABB $\rightarrow$ OBB$_\text{PC}$} & 253  & 220  & 209  & 39.3  & 32.9  & 29.6  \\ 
    & Bistro & {\tiny \textcolor{teal}{Rel. Factor to AABB}} & \tiny\textcolor{teal}{$0.48\times$} & \tiny\textcolor{teal}{$0.45\times$} & \tiny\textcolor{teal}{$0.45\times$} & \tiny\textcolor{teal}{$0.64\times$} & \tiny\textcolor{teal}{$0.63\times$} & \tiny\textcolor{teal}{$0.61\times$} \\
    & Exterior & {\tiny AABB $\rightarrow$ OBB$_\text{soCPU}$} & 256  & 225  & 213  & 40.8  & 34.2  & 30.7  \\ 
    & [2.8M] & & \tiny\textcolor{teal}{$0.49\times$} & \tiny\textcolor{teal}{$0.46\times$} & \tiny\textcolor{teal}{$0.46\times$} & \tiny\textcolor{teal}{$0.66\times$} & \tiny\textcolor{teal}{$0.65\times$} & \tiny\textcolor{teal}{$0.64\times$} \\
    & & {\tiny AABB $\rightarrow$ OBB$_\text{soGPU}$} & 361  & 317  & 316  & 52.5  & 43.5  & 39.3  \\
    & & & \tiny\textcolor{teal}{$0.69\times$} & \tiny\textcolor{teal}{$0.64\times$} & \tiny\textcolor{teal}{$0.68\times$} & \tiny\textcolor{teal}{$0.86\times$} & \tiny\textcolor{teal}{$0.83\times$} & \tiny\textcolor{teal}{$0.81\times$} \\
    \cline{2-9}
    \cline{2-9}
    \multirow{4}{*}{\includegraphics[width=0.158\textwidth]{screenshot_white_oak_mitsuba.png}} & & & & & & & &  \\
    & & \multirow{-2}{*}{\tiny AABB (Baseline)} & \multirow{-2}{*}{189} & \multirow{-2}{*}{157} & \multirow{-2}{*}{147} & \multirow{-2}{*}{15.2} & \multirow{-2}{*}{13.0} & \multirow{-2}{*}{12.0} \\  
    & & {\tiny AABB $\rightarrow$ OBB$_\text{PC}$} & 155  & 111  & 104  & 12.7  & 10.9  & 10.0  \\
    & White Oak & & \tiny\textcolor{teal}{$0.82\times$} & \tiny\textcolor{teal}{$0.71\times$} & \tiny\textcolor{teal}{$0.71\times$} & \tiny\textcolor{teal}{$0.84\times$} & \tiny\textcolor{teal}{$0.84\times$} & \tiny\textcolor{teal}{$0.84\times$} \\
    & [36K] & {\tiny AABB $\rightarrow$ OBB$_\text{soCPU}$} & 170  & 131  & 121  & 13.8  & 12.0  & 11.2  \\ 
    & & & \tiny\textcolor{teal}{$0.90\times$} & \tiny\textcolor{teal}{$0.83\times$} & \tiny\textcolor{teal}{$0.82\times$} & \tiny\textcolor{teal}{$0.91\times$} & \tiny\textcolor{teal}{$0.93\times$} & \tiny\textcolor{teal}{$0.93\times$} \\
    & & {\tiny AABB $\rightarrow$ OBB$_\text{soGPU}$}  & 184  & 146  & 128  & 14.7  & 12.6  & 11.6  \\
    & & & \tiny\textcolor{teal}{$0.97\times$} & \tiny\textcolor{teal}{$0.93\times$} & \tiny\textcolor{teal}{$0.87\times$} & \tiny\textcolor{teal}{$0.97\times$} & \tiny\textcolor{teal}{$0.97\times$} & \tiny\textcolor{teal}{$0.97\times$} \\
    \cline{2-9}
    \cline{2-9}
    \multirow{4}{*}{\includegraphics[width=0.158\textwidth]{screenshot_dragon_mitsuba.png}} & & & & & & & & \\
    & & \multirow{-2}{*}{\tiny AABB (Baseline)} & \multirow{-2}{*}{226} & \multirow{-2}{*}{195} & \multirow{-2}{*}{194} & \multirow{-2}{*}{9.0} & \multirow{-2}{*}{7.5} & \multirow{-2}{*}{6.8} \\
    & & {\tiny AABB $\rightarrow$ OBB$_\text{PC}$} & 120  & 98  & 91  & 7.6  & 6.4  & 5.8  \\
    & Dragon & & \tiny\textcolor{teal}{$0.53\times$} & \tiny\textcolor{teal}{$0.50\times$} & \tiny\textcolor{teal}{$0.47\times$} & \tiny\textcolor{teal}{$0.85\times$} & \tiny\textcolor{teal}{$0.85\times$} & \tiny\textcolor{teal}{$0.84\times$} \\
    & [830K] & {\tiny AABB $\rightarrow$ OBB$_\text{soCPU}$} & 109  & 90  & 81  & 7.8  & 6.6  & 6.0  \\
    & & & \tiny\textcolor{teal}{$0.48\times$} & \tiny\textcolor{teal}{$0.46\times$} & \tiny\textcolor{teal}{$0.42\times$} & \tiny\textcolor{teal}{$0.87\times$} & \tiny\textcolor{teal}{$0.88\times$} & \tiny\textcolor{teal}{$0.88\times$} \\
    & & {\tiny AABB $\rightarrow$ OBB$_\text{soGPU}$} & 154  & 131  & 110  & 8.6  & 7.1  & 6.5  \\
    & & & \tiny\textcolor{teal}{$0.68\times$} & \tiny\textcolor{teal}{$0.67\times$} & \tiny\textcolor{teal}{$0.57\times$} & \tiny\textcolor{teal}{$0.96\times$} & \tiny\textcolor{teal}{$0.95\times$} & \tiny\textcolor{teal}{$0.95\times$} \\
    \cline{2-9}
    \cline{2-9}
    \multirow{4}{*}{\includegraphics[width=0.158\textwidth]{screenshot_powerplant_mitsuba.png}} & & & & & & & & \\
    & & \multirow{-2}{*}{\tiny AABB (Baseline)} & \multirow{-2}{*}{2892} & \multirow{-2}{*}{2625} & \multirow{-2}{*}{2061} & \multirow{-2}{*}{31.6} & \multirow{-2}{*}{26.1} & \multirow{-2}{*}{23.2} \\
    & & {\tiny AABB $\rightarrow$ OBB$_\text{PC}$} & 447  & 467  & 475  & 25.8  & 20.9  & 18.4  \\
    & Powerplant & & \tiny\textcolor{teal}{$0.15\times$} & \tiny\textcolor{teal}{$0.18\times$} & \tiny\textcolor{teal}{$0.23\times$} & \tiny\textcolor{teal}{$0.82\times$} & \tiny\textcolor{teal}{$0.80\times$} & \tiny\textcolor{teal}{$0.79\times$} \\
    & [12.8M] & {\tiny AABB $\rightarrow$ OBB$_\text{soCPU}$} & 510  & 471  & 425  & 26.8  & 21.6  & 18.9  \\
    & & & \tiny\textcolor{teal}{$0.18\times$} & \tiny\textcolor{teal}{$0.18\times$} & \tiny\textcolor{teal}{$0.21\times$} & \tiny\textcolor{teal}{$0.85\times$} & \tiny\textcolor{teal}{$0.83\times$} & \tiny\textcolor{teal}{$0.82\times$} \\
    & & {\tiny AABB $\rightarrow$ OBB$_\text{soGPU}$} & 1573  & 1358  & 971  & 27.4  & 22.3  & 19.6  \\
    & & & \tiny\textcolor{teal}{$0.54\times$} & \tiny\textcolor{teal}{$0.52\times$} & \tiny\textcolor{teal}{$0.47\times$} & \tiny\textcolor{teal}{$0.87\times$} & \tiny\textcolor{teal}{$0.85\times$} & \tiny\textcolor{teal}{$0.84\times$} \\
    \cline{2-9}
    \cline{2-9}
  \end{tabular}
  \caption{BVH quality and traversal statistics for secondary (GI) rays across different BVH widths for all datasets and OBB conversion algorithms.}
  \label{tab:data_sah_iters_gi}
\end{table*}

{
\setlength{\tabcolsep}{0.5pt}
\begin{table*}[!h]
\centering
\begin{tabular}{cccl}
\cellcolor{darkgray}\textcolor{white}{AABB BVH-8} & \cellcolor{darkgray}\textcolor{white}{$\rightarrow$ OBB$_\text{soCPU}$ (Brute-Force)} &  \cellcolor{darkgray}\textcolor{white}{$\rightarrow$ OBB$_\text{soGPU}$ (Our Approach)} & \\
\includegraphics[width=.28\textwidth]{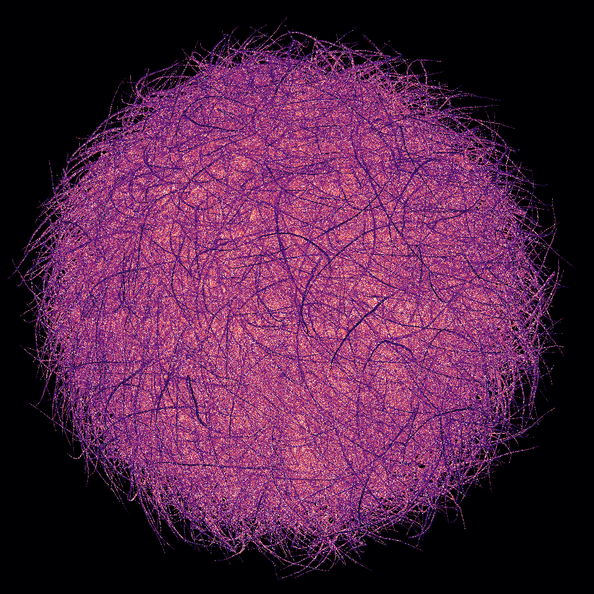} &
\includegraphics[width=.28\textwidth]{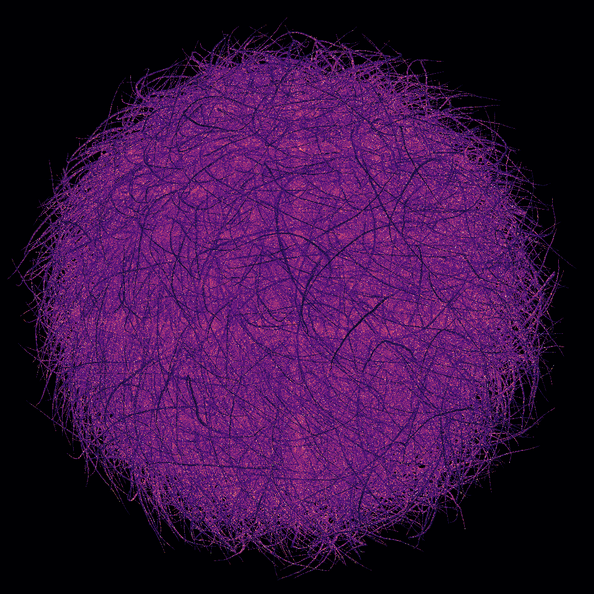} &
\includegraphics[width=.28\textwidth]{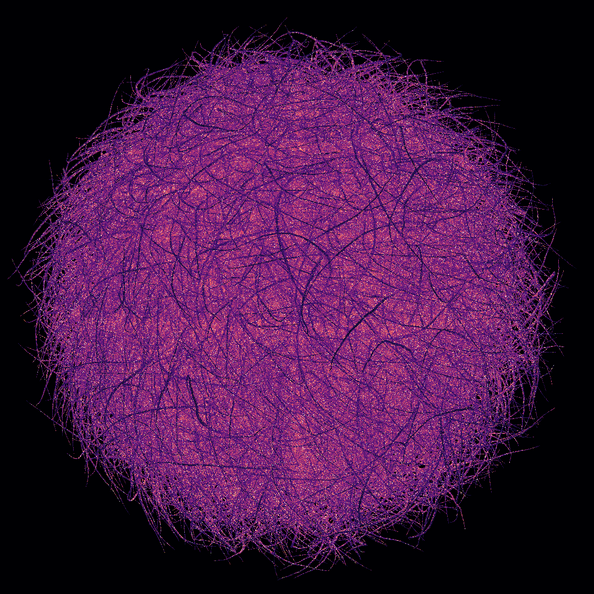} & \includegraphics[height=0.2\textheight]{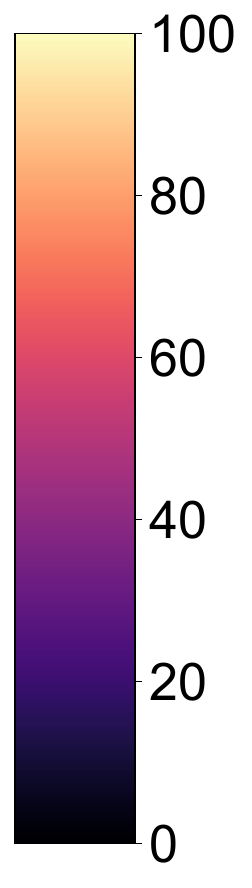} \\ [-2pt]
\includegraphics[width=.28\textwidth]{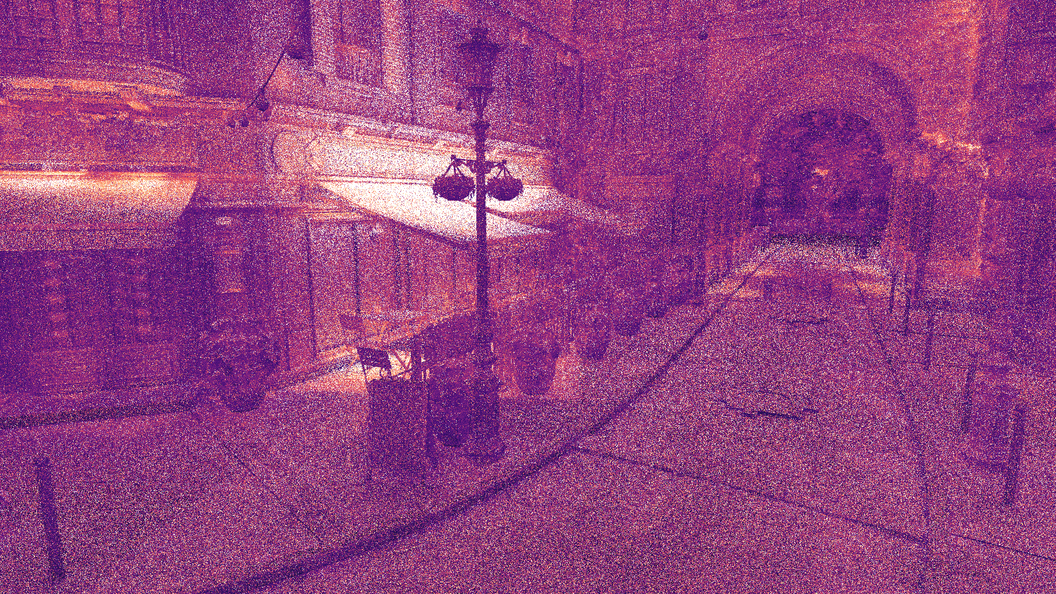} &
\includegraphics[width=.28\textwidth]{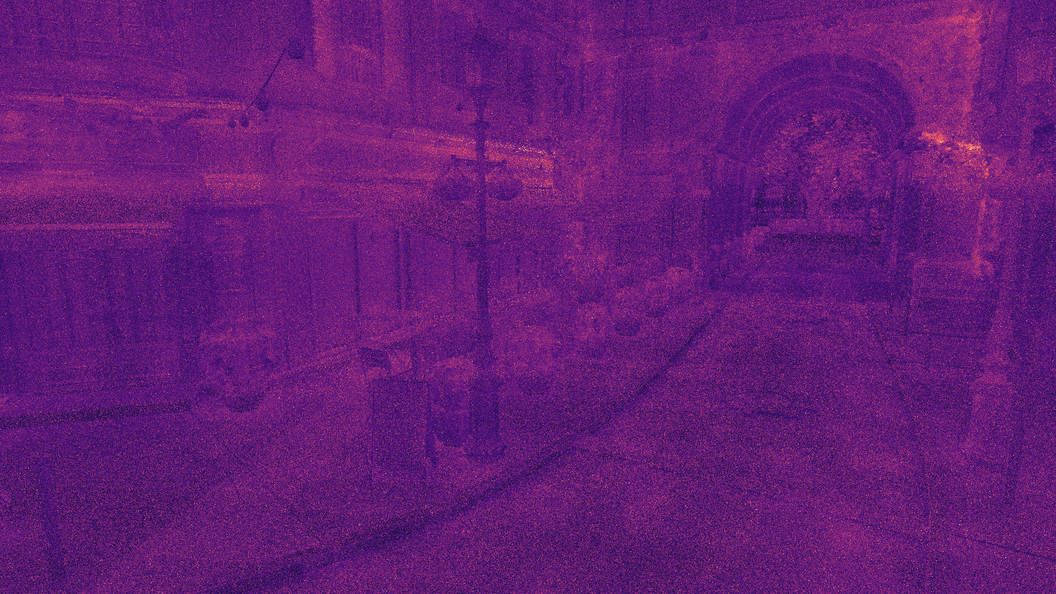} &
\includegraphics[width=.28\textwidth]{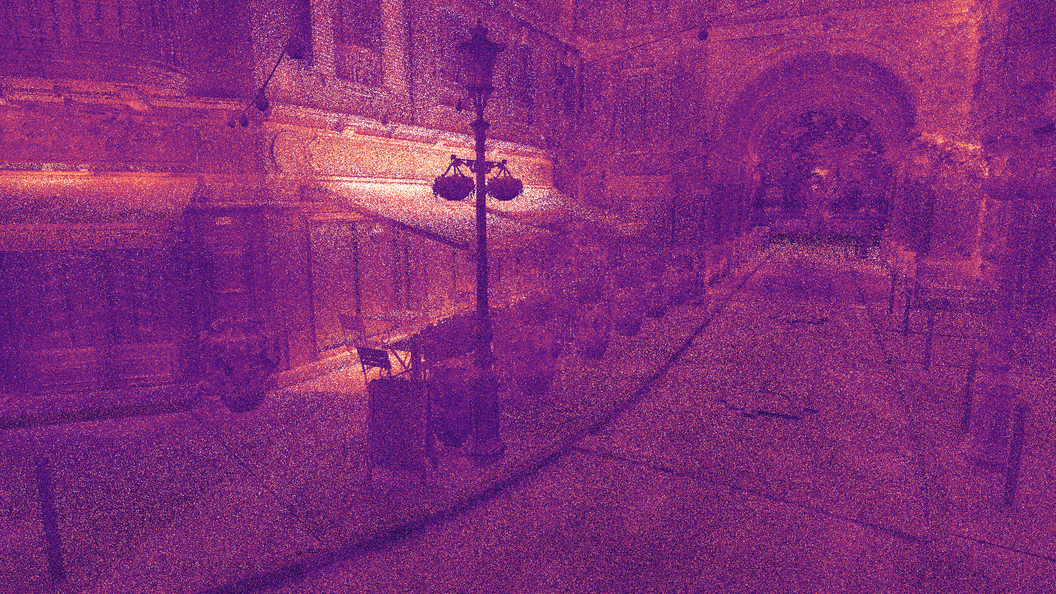} & \includegraphics[height=0.12\textheight]{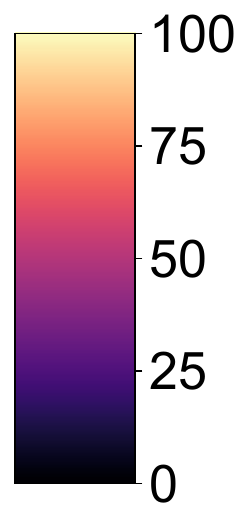} \\ [-2pt]
\includegraphics[width=.28\textwidth]{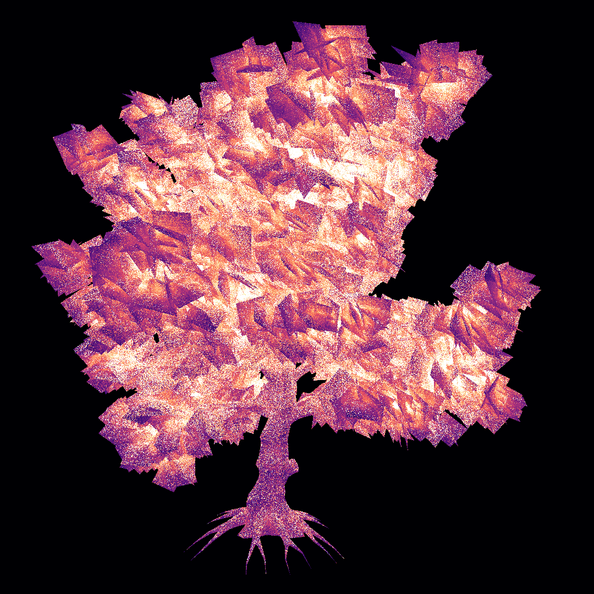} &
\includegraphics[width=.28\textwidth]{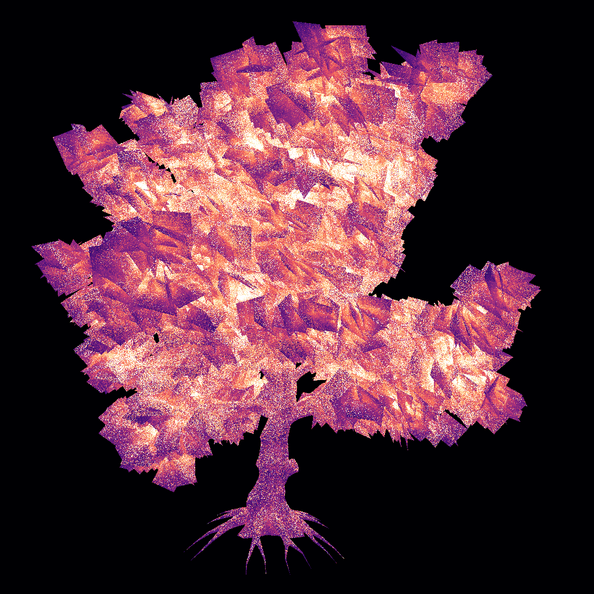} &
\includegraphics[width=.28\textwidth]{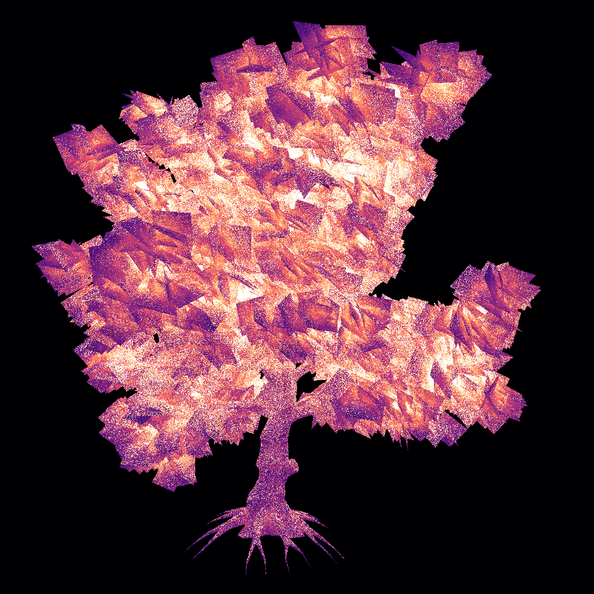}  & \includegraphics[height=0.2\textheight]{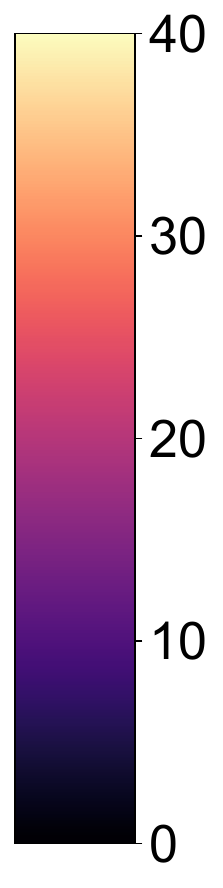} \\ [-2pt]
\includegraphics[width=.28\textwidth]{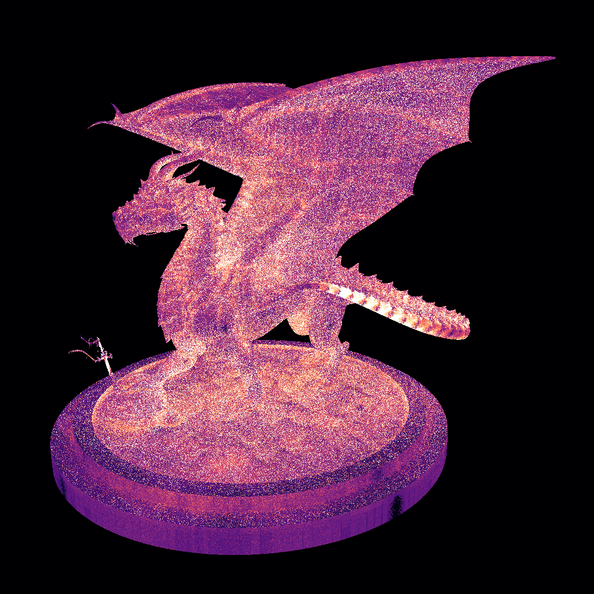} &
\includegraphics[width=.28\textwidth]{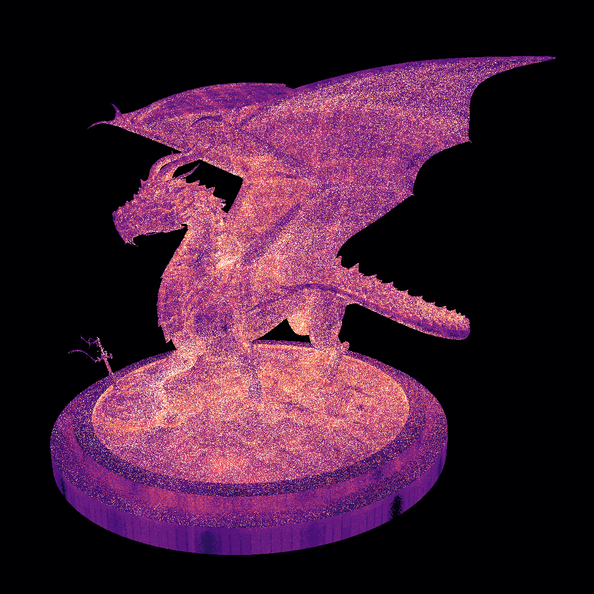} &
\includegraphics[width=.28\textwidth]{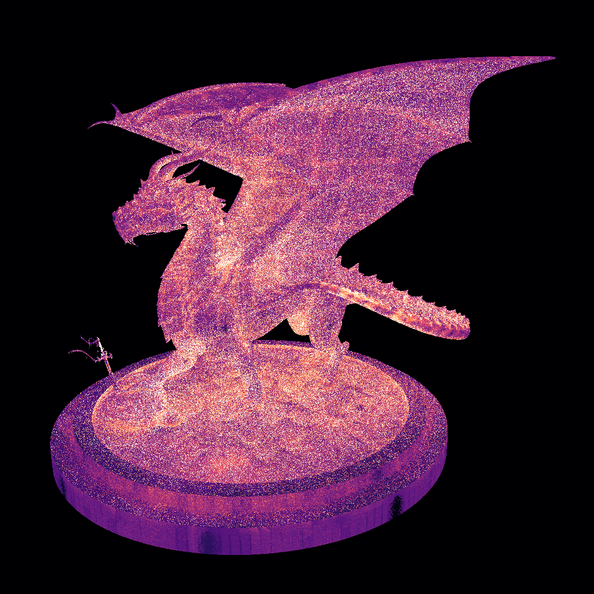}  & \includegraphics[height=0.2\textheight]{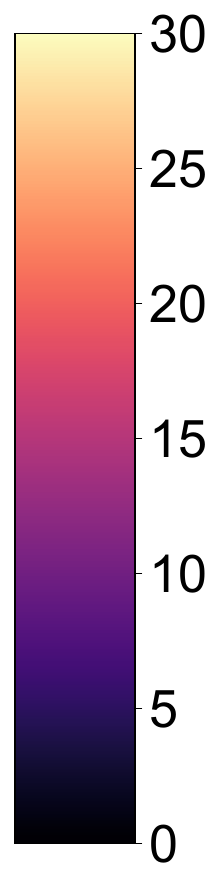} \\ [-2pt]
\includegraphics[width=.28\textwidth]{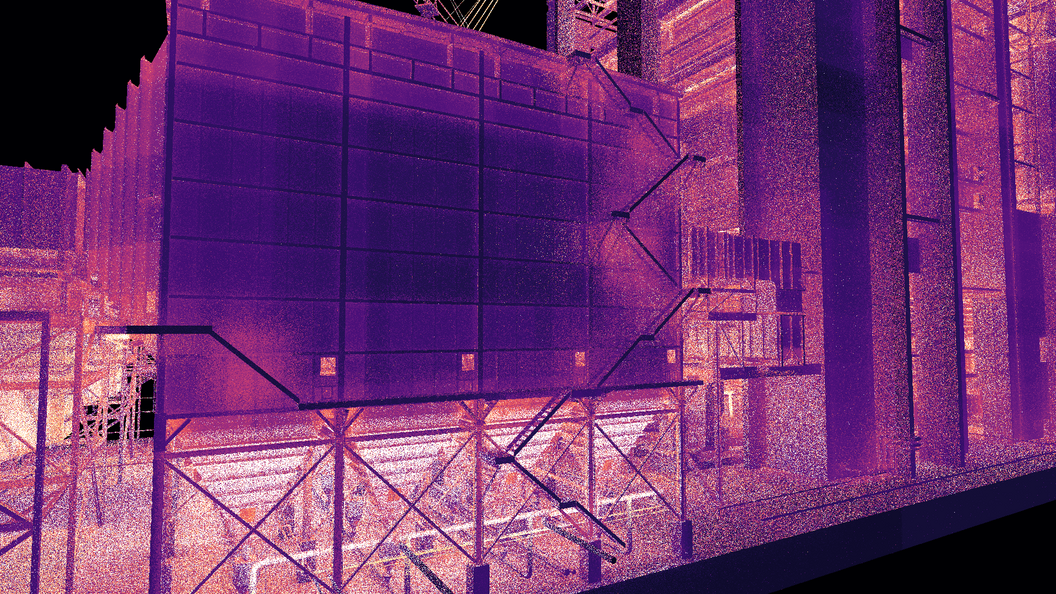} &
\includegraphics[width=.28\textwidth]{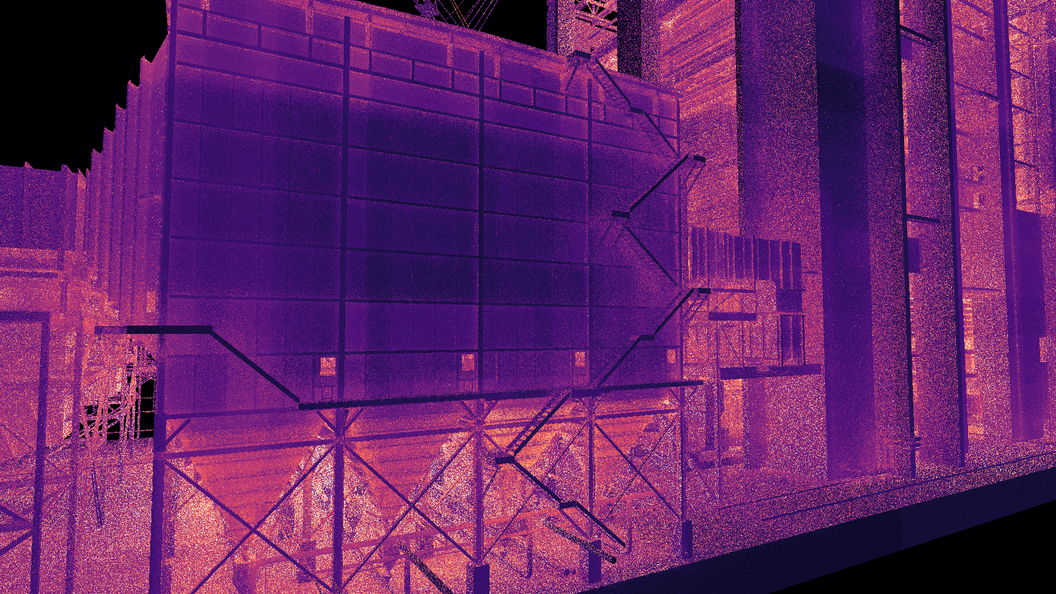} &
\includegraphics[width=.28\textwidth]{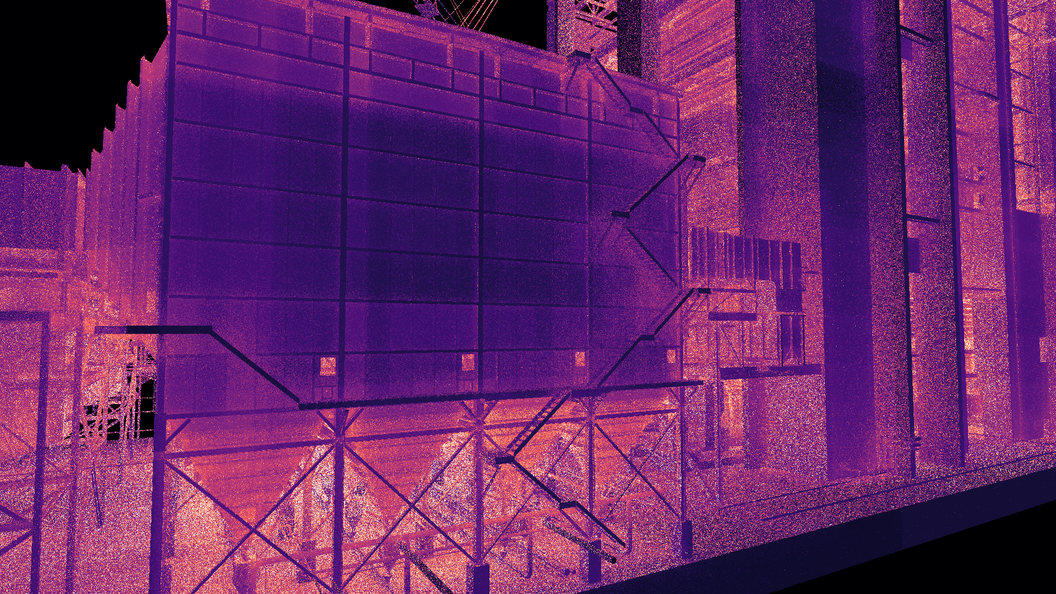} & \includegraphics[height=0.12\textheight]{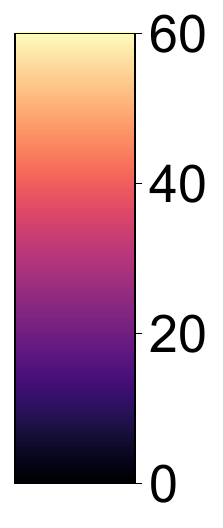} \\
\end{tabular}
\caption{Visualizaton of traversal loop iteration counts at each pixel for secondary (GI) rays for our CPU-based and GPU-based DOBB BVHs compared to the baseline AABB BVH.}
\label{tab:bvh8_iters_gi}
\end{table*}
}

\end{document}